**Title:** Noise-induced oscillatory shuttling of NF-κB in a two compartment IKK-NF-κB-IκB-A20 signaling model

**Authors:** Jaewook Joo[1,*], Steven J. Plimpton[2], Jean-Loup Faulon[3]

**Affiliations:** [1]Department of Physics and Astronomy, University of Tennessee, Knoxville, 37996, USA, [2]Scalable Algorithms Department, Sandia National Laboratories, Albuquerque, NM, 87185, USA; [3]Department of Biology, Every University, Every Cedex, France.

*Corresponding author's email: jjoo1@utk.edu

## ABSTRACT

NF-κB is a pleiotropic protein whose nucleo-cytoplasmic trafficking is tightly regulated by multiple negative feedback loops embedded in the NF-κB signaling network and contributes to diverse gene expression profiles important in immune cell differentiation, cell apoptosis, and innate immunity. The intracellular signaling processes and their control mechanisms, however, are susceptible to both extrinsic and intrinsic noise. By extrinsic noise we mean heterogeneous intracellular physiological conditions resulting in variations in kinetic rate constants between single cells. Intrinsic noise originates from the probabilistic nature of intracellular biochemical reactions. This poses an important question, "how do extrinsic and intrinsic noise affect the nucleo-cytoplasmic shuttling dynamics of NF-κB, which ultimately leads to variations in single cell response?" In this article, we present numerical evidence for a universal dynamic behavior of NF-κB, namely oscillatory nucleo-cytoplasmic shuttling, due to the fundamentally stochastic nature of the NF-κB signaling network. We simulated the effect of extrinsic noise with a deterministic ODE model, using a statistical ensemble approach, generating many copies of the signaling network with different kinetic rates sampled from a biologically feasible parameter space. We modeled the effect of intrinsic noise by simulating the same networks stochastically using the Gillespie algorithm. The results demonstrate that extrinsic noise diversifies the shuttling patterns of NF-κB response, whereas intrinsic noise induces oscillatory behavior in many of the otherwise non-oscillatory patterns. We identify two key model parameters which significantly affect the NF-κB dynamic response and deduce a two-dimensional phase-diagram of the NF-κB response as a function of these parameters. We conclude that if single-cell experiments are performed, a rich variety of NF-κB response will be observed, even if population-level experiments, which average response over large numbers of cells, do not evidence oscillatory behavior.

## AUTHOR SUMMARY

Cells respond to external stress in a regulated manner, using negative feedback loops to tightly control the nuclear concentration of activators that can enhance the expression of genes that help the cell survive. A major unresolved question is how individual cells make critical cell-fate decisions in the presence of stochastic fluctuations in intracellular biochemical reactions. We attempt to answer this question with computations on a model of the NF-κB signaling network responsible for immune response, apoptosis, and cell growth. To this point there has been no consensus about oscillatory shuttling of NF-κB between nucleus and cytoplasm. Relying on computational analyses, we describe the underlying cellular mechanisms responsible for oscillations of NF-κB. We show that the interplay of intracellular stochastic fluctuations and negative feedback loops can actually enhance the likelihood of an oscillatory response by NF-κB within a biologically feasible parameter domain. We also argue that this effect will be most easily observed in single-cell experiments and may be masked in population level measurements due to the heterogeneous nature of individual cell responses. This result may be widely applicable to other similar cell signaling systems, meaning oscillations in the concentration of key proteins may be a more common response than previously thought.

## I. INTRODUCTION

Intracellular biochemical processes can be extremely noisy because the copy number of crucial proteins is small, roughly in order of as low as tens for prokaryotic cells and hundreds for eukaryotic cells [1-7]. Cells, however, manage to detect, transmit, and process external signals robustly, and make crucial decisions consistently despite noisy intracellular information processing machineries. A great puzzle in "quantitative biology" is how cells produce precise and robust gene expression profiles by using noisy control feedback mechanisms embedded in signal transduction networks or gene-regulatory networks. Ground breaking work has been performed to enhance our understanding of the effects of intrinsic noise on the performance of gene regulatory networks [8-17]. However, only a very few signal transduction networks have been analyzed this way due to lack of detailed information about network structure and kinetics. In this paper we address the effects of both extrinsic and intrinsic noise on the dynamic response of key proteins in an intracellular signal transduction network, emphasizing intrinsic noise-induced oscillation and extrinsic noise-diversified dynamical response.

The oscillation in concentration of key proteins is one of the most ubiquitously observed responses in regulatory and signaling pathways in cells because of negative feedback loops where a gene eventually inhibits its own transcription through one or more intermediate transcribed proteins. This negative feedback loop motif is found far more often than would be predicted by chance in various species [18-20]. In addition to well-known examples such as circadian rhythm [21] and cell cycles [22], many cellular signaling systems exhibit pronounced oscillations of key proteins such as p53 [23], Hes1 [24, 25] and NF-κB [26, 27]. For the case of p53 oscillations, there is a dramatic difference between population-level and single-cell studies of p53 expression in response to DNA damage; here a network feedback loop consists of p53 and Mdm2 protein. The latter's transcription is activated by p53 and causes p53 degradation [28]. Population level studies show a damped oscillatory response of p53 [29] while, in contrast, single-cell studies exhibit sustained oscillations of p53 [23]. Surprisingly, genetically identical cells exposed to the same amount of DNA-damaging radiation respond with a variable number of p53 pulses at fixed time duration [23, 30]. A similar difference between population level and single-cell level studies is also observed for the nucleo-cytoplasmic shuttling of NF-κB in response to TNFα stimulation [26, 27] and is discussed in detail below.

Ref. [31] proposed that the oscillations of p53 as well as those of NF-κB could have essential cellular functions: a refresh process to respond to a new signal, or a cellular protection mechanism from the high dosage-induced damage. Both NF-κB and p53 are key players in making crucial decisions related to cell fate. If uncontrolled and over-expressed, high concentrations of these proteins can be fatal to the host organism [Lee et al 2000]. Thus they are tightly regulated by negative feedback loops, resulting in oscillations as a dynamical cellular response in attempts to balance the competing outcomes of maximal protein dosage and minimal host damage.

Despite these insights, the existence of, the cellular mechanisms behind, and the biological function of oscillatory shuttling of NF-κB are largely unresolved questions and have been the subject of recent debates [27, 32-34]. NF-κB is a transcription factor that regulates the expression of numerous genes in response to UV, Tumor Necrosis Factor, Lipopolysaccharide, and antigens stimulations, with roles in cellular stress responses, cell growth, apoptosis, and immune response [35-37]. The negative feedback loops in the NF-κB signaling network consist of NF-κB-dependent IκBα, IκBε, A20 genes and the constitutive IκBβ gene, whose protein products sequester NF-κB in the cytoplasm and can generate oscillatory nucleo-cytoplasmic shuttling of NF-κB. The population level studies in Ref. [26] showed damped oscillatory NF-κB shuttling for a wild type and a mutant with IκBβ and IκBε genes knocked out and showed non-oscillatory shuttling for other double mutants of IκB isoform genes. Another population level study in Ref. [38, 39] showed no oscillatory NF-κB shuttling when the A20 gene was knocked out. Using single-cell level study, Ref. [27] reports noisy quasi-oscillatory NF-κB shuttling but only in a fraction of TNFα-stimulated cells (30% of human cervical carcinoma cells and 70% of human S-type neuroblastoma cells). The observed oscillations of NF-κB in this single-cell level study can be viewed as an abnormal behavior of the minority of cells. The experimental results in Ref. [27] are in the presence of multifold increases in RELA and IκBα protein copy numbers because of artificial insertion of two RELA and IκBα reporter genes into the DNA [32]. These two points cast a reasonable doubt on the existence of sustained oscillations of NF-κB in wild type and individual cells. More importantly, Ref. [32] argues that both non-oscillatory and oscillatory NF-κB shuttling yields the same gene expression profiles, which disputes a role for oscillations of NF-κB on gene transcription. The findings in Ref. [32] are consistent with an increasing number of papers emphasizing the essential roles of signal-induced histone modifications and chromatin remodeling as an additional layer of transcriptional regulation [40-44].

Various computational models of the NF-κB signaling network have been proposed to elucidate the underlying mechanisms of the NF-κB shuttling dynamics. In Ref. [26], a computational ODE model identified distinctive roles for each of three isoforms of IκB whereas Refs. [45] and [46] emphasized the effect of the temporal profile of IKK activity on NF-κB dynamics [26, 45, 46]. In Ref. [39], this ODE model was enhanced to include the effects of the volume ratio of cytoplasm to nucleus on the concentration of proteins shuttling between the nucleus and cytoplasm, and an additional negative regulator A20 was added which inactivates IKK [38]. Ref. [47] numerically analyzed the reduced order ODE model of an IκBβ/ε -/- double mutant case (consisting of nuclear NF-κB, cytoplasmic IκBα, and IκBα mRNA) and demonstrated a very robust spiky oscillatory shuttling of NF-κB. The saturated degradation of IκBα was argued as the origin of this dynamical robustness. Additionally, the stochastic effects on the NF-κB shuttling dynamics were investigated to show how damped-oscillatory dynamics of NF-κB at the population level can arise from the ensemble average of noisy damped-oscillatory shuttling over many realizations of identical networks [48, 49].

In this paper we argue that all of the above seemingly disparate observations and arguments about NF-κB oscillations can be reconciled in one consistent story. This is motivated by three observations. First, we note that the dynamic response of the NF-κB signaling network to the same dosage of a stimulant can be dramatically different among individual cells, even derived from the same parental cells as single-cell level studies of both p53 and NF-κB alluded to [23, 27, 30, 31]. Differences in the phase of the cell cycle, the basal level of protein expression, or any other stochastic variation can contribute to cell-to-cell variations in the NF-κB response. Therefore, only single-cell level analyses, either experimentally or computationally, can elucidate actual control mechanisms responsible for diverse responses. Population level analyses whose measurements are averaged over a population of cells can mask the individuality and heterogeneity of the cells. Second, the NF-κB-induced gene expression profiles may result from the dynamic interplay of equally-important functional modules such as the signaling pathway that regulates NF-κB nucleo-cytoplasmic shuttling, the signaling network of chromatin remodeling that controls the accessibility of activators to the DNA site, and the signaling pathways of other activators cooperatively required for NF-κB-dependent gene transcription [40-44]. In the framework of “modular” cell biology [50], we consider the NF-κB-induced gene expression machinery to be an integrated system consisting of multiple functional modules, each of which consists of interacting molecules with a function that is separable from those of other modules. Naturally a full understanding of the dynamics of individual modules should come before a comprehensive understanding of NF-κB-induced gene expression profiles as an output of the integrated system. Thirdly, a computational model should be able to test the hypothesis of oscillations of NF-κB in the wild type cells at the level of single cells. Current single-cell experimental methods rely on reporter gene constructs, inevitably mutating the cells, if one is to track the spatio-temporal dynamics of a key protein within single cells. From these perspectives, we revisit computational investigations of the spatio-temporal dynamics of NF-κB protein at the level of single cells, taking into account both extrinsic and intrinsic stochastic variations between individual cells.

In this paper, we investigate how the shuttling dynamics of NF-κB is regulated in a noisy intracellular environment at the level of single-cells. For this purpose, we construct both deterministic and stochastic two compartment models (cytoplasm vs. nucleus) of the NF-κB signaling network and study the effects of both extrinsic and intrinsic noise on the NF-κB dynamic response. Simulation of extrinsic noise is performed by randomly sampling of kinetic rate values from a biologically feasible domain. We see that such noise enriches the diversity of NF-κB shuttling responses. The responses can be classified into well-characterized patterns, only a small fraction of which are oscillatory. We then show that intrinsic noise (intracellular stochastic fluctuations) can induce oscillatory shuttling of NF-κB even in regions of the biologically feasible parameter domain that yield only non-oscillatory response in the deterministic model. We also map a phase-diagram of the noise-induced oscillation and other dynamic NF-κB patterns as a function of two key model parameters, which were found to be most significant by sensitivity analyses. We also demonstrate that both linear and saturating transcription models produce similar phase diagrams. We find the saturating model suppresses oscillations somewhat, but intrinsic noise enhances them, giving rise to overall oscillatory behavior in a significant fraction of biologically feasible parameter space.

## II. RESULTS

We characterize the shuttling dynamics of *NF-κB* in a noisy intracellular environment in the following results sections. In section II-A, we use a simple transcription model where transcription activity is linearly dependent on nuclear *NF-κB* concentration. With extrinsic noise and a deterministic ODE model of the *NF-κB* signaling network, we find only a small fraction of the biologically feasible kinetic rates yield an oscillatory response and plot a two-variable bifurcation phase diagram as a function of two key parameters of the *NF-κB* signaling network: the total amount of *NF-κB* protein ($[NF\kappa B_o]$) and the volume ratio of cytoplasm to nucleus ($Kv$). Then, stochastic simulations of the same model with the same parameter values show that the probability of oscillatory *NF-κB* shuttling is greatly enhanced by inclusion of intracellular stochastic fluctuations. This is indicated by an increased area of oscillatory response within the phase diagram for a fixed set of kinetic rates and by an increased probability of an oscillatory response when the rates are chosen randomly. In section II-B, we present results from a modified transcription model where transcription activity saturates once nuclear *NF-κB* concentration reaches a threshold level.

**A. Linear Transcription model: Deterministic Case**

We start with an assumption of linear transcription, a commonly used model in various literatures [26, 27, 39], where the rate of mRNA synthesis increases linearly with *NF-κB* concentration in the nucleus. We regard the dynamic patterns of nuclear *NF-κB* concentration as the single most important output of the *NF-κB* signaling network shown in Fig. 1 in response to a persistent stimulation. Using Latin Hypercube Sampling (LHS) techniques (see methods), we generated a thousand different sets of kinetic rates for our model. Each of the 71 parameters (mostly rates) was allowed to vary over an interval (*0.3Xi, 1.7Xi*) where *Xi* denotes its nominal value. Using these thousand sets as a statistical ensemble of the network, we generated a thousand "possible" and inhomogeneous temporal profiles of nuclear *NF-κB* concentration. Following Ref. [51], we use these heterogeneous *NF-κB* responses in two different ways: first, for sensitivity analysis, and then as representative responses of individual cells with phenotypic variations. For sensitivity analysis, we measured 5 metrics for each *NF-κB* profile: a *Steady State* value from the asymptotic level of *NF-κB* concentration, a *Period between oscillatory peaks, a Damping Constant* computed from the decrease in amplitude between the first and second oscillatory peaks, a *Phase* as the delay time between stimulation and the first peak of *NF-κB* concentration, and a *First Max value* as the height of the first peak. Refs. [39, 51] showed that these 5 metrics were most highly correlated with two input parameters, the total *NF-κB* concentration ($[NF\kappa B_o]$) and the volume ratio of cytoplasm to nucleus (*Kv*). In Fig. 2 the correlation between each of the 5 metrics and $[NF\kappa B_o]$ or *Kv* are presented. Both *First Max* and *Damping Constant* are positively correlated with $[NF\kappa B_o]$ and *Kv* while both *Period and Phase* are negatively correlated; *First Max* and *Damping Constant* increase and *Period and Phase* decrease when either $[NF\kappa B_o]$ or *Kv* increases. In other words, the *NF-κB* response, represented by a temporal profile of nuclear *NF-κB* concentration, is bigger (larger first max), faster (smaller phase delay), and more dynamically modulated (larger damping constant and shorter period) for larger values of $[NF\kappa B_o]$ and *Kv*. This observation leads us to hypothesize that oscillatory shuttling of *NF-κB* can be experimentally observed for large values of $[NF\kappa B_o]$ and *Kv*, as the single cell level.

To validate the conjecture that the likelihood of NF-κB oscillatory shuttling increases with the increase of either $[NF\kappa B_o]$ or *Kv*, we used the deterministic ODE model of the NF-κB network with one thousand sets of kinetic parameters sampled in two ways (either small or large values of $[NF\kappa B_o]$ or *Kv*) and calculated the fraction of the oscillatory response for each case. In the first set of thousand inputs, $[NF\kappa B_o]$ and *Kv* were LHS-sampled from the intervals, (*0.01 μM,* 0.1 μM) and (1, 10) respectively. In the second set, $[NF\kappa B_o]$ and *Kv* were fixed at the largest value in those intervals, say, 0.1 μM and 10, respectively. In both sets of inputs, the remaining 69 input parameter values were LHS-sampled as before, from a span of biologically feasible values. With these two types of inputs, the resulting distribution of *NF-κB* shuttling response changes dramatically, as shown in Fig. 3. For the former set of inputs, 2% of the *NF-κB* profiles have sustained-oscillations, 7% are single-peaked, 13% are monotonically saturating (typical response of an over-damped system), and 78% have damped oscillations (typical of an under-damped system). (See Ref. [51] for illustrations of the above patterns.) For the second set of inputs, the fractions are 20% sustained oscillations and 80% damped oscillations. Thus the likelihood of an oscillatory *NF-κB* response increases with increasing values of $[NF\kappa B_o]$ and *Kv*.

As $[NF\kappa B_o]$ and *Kv* change, at least three different transitions between *NF-κB* dynamic response patterns can take place. All possible transitions as a function of $[NF\kappa B_o]$ and *Kv* are presented in Table II. About 20% of the LHS-sampled parameter sets exhibit a transition from damped-oscillatory to sustained-oscillatory *NF-κB* response as $[NF\kappa B_o]$ and *Kv* increase from small to largest values. About 10% show two other transitions, either from single-peaked response to sustained-oscillatory response, or from monotonically saturating response to damped-oscillatory response.

We now focus on the transition from a damped-oscillatory pattern to a sustained-oscillatory pattern (i.e., one with a dynamical instability or limit cycle), as a function of $[NF\kappa B_o]$ and *Kv*. In Fig. 4 we show a bifurcation phase diagram of the *NF-κB* signaling network as a function of two parameters $[NF\kappa B_o]$ and *Kv*. The remaining 69 input parameters were chosen from one of the input data sets generated by the LHS procedure, for a case that generated sustained-oscillatory shuttling of *NF-κB* for the largest values of $[NF\kappa B_o]$=0.1μM and *Kv=10.* Holding these 69 parameters fixed, $[NF\kappa B_o]$ was then varied from 0.01 to 0.1 μM and *Kv* from 1 to 10 and the *NF-κB* responses were classified as either sustained-oscillatory or damped-oscillatory. Fig. 4 illustrates the phase boundary line separating the sustained-oscillatory (OSC) domain from the damped-oscillatory (DAMP) domain as well as the examples of the sustained-oscillatory and the damped-oscillatory temporal profiles of nuclear *NF-κB* concentration at two different values of $[NF\kappa B_o]$ and *Kv*. Other bifurcation diagrams, obtained from different fixed settings of the 69 remaining value gave similar results, though the precise position of the phase boundary can vary from case to case. Similar phase diagrams for transitions from monotonically saturating response to damped-oscillatory response or from single-peaked response to damped-oscillatory response can also be constructed.

### B. Linear Transcription Model: Stochastic Case

We now investigate the effects of intrinsic noise on the *NF-κB* response, especially on the distribution of the *NF-κB* dynamic patterns shown in Fig. 3 and the bifurcation diagram shown in Fig. 4. We performed Gillespie simulations [52] of the same reaction network using the same sets of input rate values that generated the bifurcation diagram of Fig. 4. We set the cell volume (cytoplasm plus nucleus) to 2000 $\mu m^3$. Both this volume and the volume ratio (*Kv*) determine the propensity of reactions taking place in the cytoplasm or in the nucleus in the Gillespie model. Note that a typical *NF-κB* concentration of 0.05 *μM* becomes a discrete molecule count of about 60,000 in our model cell. At the highest values of $[NF\kappa B_o]=0.1\ \mu M$ and *Kv=9*, the stochastic temporal profile of nuclear *NF-κB* concentration is very similar to the deterministic result as shown in Fig. 5(a). This indicates that this deterministic result is robust even with stochastic fluctuations. At intermediate values with $[NF\kappa B_o]=0.06\ \mu M$ and *Kv=4,* where the deterministic model yields a stable fixed point, the stochastic profile in Fig. 5(c) exhibits a noisy, yet prominent, oscillation. At small values of $[NF\kappa B_o]=0.04\ \mu M$ and *Kv=2,* the fluctuations in the stochastic time-series data in Fig. 5(e) are of the same magnitude the nuclear *NF-κB* concentration and mask whatever signal is present.

As illustrated in Fig. 5(a), 5(c), and 5(e), the stochastic time-series data are too noisy to definitively classify as oscillatory or not by our previous scoring metrics. Instead we used Fourier analysis and inspected the resulting power spectra; see Fig. 5(b), 5(d), and 5(f) for the power spectra of Fig. 5(a), 5(c), and 5(e), respectively. We introduce two criteria for noise-induced oscillation: (a) occurrence of the power spectrum peak at non-zero frequency and (b) a Signal to Noise Ratio (*SNR)* greater than one where the *SNR* is defined as the ratio of the peak amplitude of a power spectrum to $\sqrt{[NF\kappa B_0]}$, representing a typical fluctuation of protein concentration in the network (see methods). All the spectra of Fig. 5 have a prominent peak at non-zero frequency and the associated *SNR* values decrease gradually from hundreds in Fig. 5(b) to smaller than one in Fig. 5(f). Thus the power spectrum of Fig. 5(f) is not classified as noise-induced oscillations while the spectra in Fig. 5(b) and 5(d) are. Using the above criteria for an oscillatory response, we present a similar phase diagram as before of stochastic *NF-κB* response shown in Fig. 6. The solid line in Fig. 6 denotes the deterministic bifurcation boundary whereas the dashed line separates the stochastic noise-induced oscillatory domain from the non-oscillatory domain. In the deterministic case, the dynamical instability domain (*OSC*) occupies about 20 % of the biologically feasible domain of $[NF\kappa B_o]$ and *Kv*. However, intrinsic noise expands the oscillatory domain to about 60% (including both *NIO* and *OSC*), i.e. a three-fold increase in the likelihood of oscillatory shuttling.

The noise-induced oscillatory shuttling of *NF-κB* observed in this study is a precursor of nearby dynamic instability as described in Ref. [55]. As clearly demonstrated in Fig. 7(a), the deterministic model generates a phase trajectory of a limit cycle at the highest values of $[NF\kappa B_o]$ and *Kv.* As they decrease to the intermediate values of $[NF\kappa B_o]$ and *Kv,* the deterministic system has a stable fixed point, but due to the presence of nearby dynamic instability (i.e., a limit cycle), the phase trajectory takes an unusually long excursion as shown in Fig, 7(b). Small stochastic fluctuations can perturb the system and the system takes a long path back to its stable fixed point. Therefore, the noise-induced oscillatory shuttling of *NF-κB* in this particular computational model is caused by continual stochastic perturbations followed by long excursions back to stable fixed points. However, at very low values of $[NF\kappa B_o]$ and *Kv,* which are far from the dynamical instability domain, the deterministic system doesn't exhibit a required long trajectory for noise-induced oscillation as shown in Fig. 7(c).

Lastly, we investigated the effects of intrinsic noise on the distribution of the *NF-κB* response patterns at the highest values of $[NF\kappa B_o]$ and *Kv.* We performed Gillespie simulations for the same set of a thousand kinetic rate inputs that generated the deterministic distribution of responses for the highest values of $[NF\kappa B_o]$=0.1 $\mu M$ and *Kv=10* shown in Fig. 3. To our surprise, almost all of the stochastic *NF-κB* profiles exhibited noise-induced oscillations. This result strongly indicates the existence of similar phase diagrams as shown in Fig. 4 for all of the thousand kinetic rate inputs, i.e., all the biologically feasible kinetic rates. Thus, oscillatory *NF-κB* shuttling is a universal feature of the *NF-κB* signaling network in single cells. Therefore, for the linear transcriptional model, we conclude that intrinsic stochastic noise induces oscillations that override the diversity of *NF-κB* dynamic patterns resulting from extrinsic noise, especially at the highest values of $[NF\kappa B_o]$=0.1 $\mu M$ and *Kv=10.*

**C. Saturating Transcription Model: Deterministic Case**

We next converted the somewhat unphysical linear transcription model used in sections II-A and II-B to a saturating transcription model and investigated the effects of transcriptional saturation on oscillatory shuttling of *NF-κB*. The motivation for the saturated model is that DNA has a finite number of binding sites for *NF-κB* proteins. When all sites are bound, the DNA is no longer responsive to an increasing *NF-κB* nuclear concentration. This effect can be modeled by having transcription reach a maximum rate once the *NF-κB* concentration passes a specified threshold value. I*n* principle the switching of individual DNA binding sites between occupied and unoccupied is stochastic, which is a dominant source of stochastic gene expressions [1-7]. Here we use a saturating transcription model where mRNA synthesis rates of *A20*, I*κB*$\alpha$*,* and *IκB*$\varepsilon$ linearly increase with small nuclear *NF-κB* concentration, but saturate to maximum values for large nuclear *NF-κB* concentration (see methods for details). The resulting transcription will be either deterministic or stochastic depending on whether the ODE or Gillespie model is used.

Using the deterministic ODE version of the *NF-κB* signaling network with the saturating transcription model, we computed distributions of the *NF-κB* response for two different choices of $[NF\kappa B_o]$ and *Kv* respectively. The results are shown in Fig. 8. On the left side (a), $[NF\kappa B_o]$ and *Kv* were randomly LHS-sampled from the intervals, (*0.01 μM,* 0.1 μM) and (1, 10) respectively and; on the right side (b), they were fixed at their largest values in these intervals. As before, the remaining 69 input parameters were LHS-sampled from biologically feasible intervals and ensembles of a thousand sets of input values were computed. In each case (a) and (b), several values of the *HM* parameter (defined as the ratio of dissociation rate to association rate between *NF-κB* and DNA) were used: *HM*=20 *nM,* 100 *nM,* and 1000 *nM*. The distribution of *NF-κB* responses clearly depends on both the value of *HM* and the values of $[NF\kappa B_o]$ and *Kv*. When the value of *HM* increases from 20 *nM* to 1000 *nM*, the onset of transcriptional saturation occurs at a higher dosage of *NF-κB*; effectively the saturating transcription model becomes a linear model across the range of biologically feasible *NF-κB* concentrations. For the case of Fig. 8 (a) and this increase of *HM* (from 20 *nM* to 1000 *nM*), the high percentage (about 50%) of single-peaked patterns dramatically decreases to less than 1%. Similarly, the percentage of sustained-oscillatory responses increase from almost nothing to about 9%. Fig. 8 also demonstrates that for a fixed value of *HM*, regardless of its value, the percentages of sustained-oscillatory and single-peaked dynamic patterns increases, respectively, as the values of $[NF\kappa B_o]$ and *Kv* are changed from (a) to (b); the similar trend was also observed in Fig. 3. These observations indicate the effect of the transcriptional saturation on the response dynamics of *NF-κB*: it suppresses oscillatory *NF-κB* shuttling.

Changing the values of $[NF\kappa B_o]$ and *Kv* while holding the value of *HM* fixed, yields two different transitions for *NF-κB* response, as presented in Table II. The first transition is between damped-oscillatory and sustained-oscillatory patterns of *NF-κB* shuttling. For a fixed value of *HM* equaling *100 nM,* almost 9 % of the damped-oscillatory patterns undergo a transition to the sustained-oscillatory pattern when *Kv* changes from (a) to (b). Fig. 10 presents a bifurcation diagram for the saturating transcription model with *HM=100 nM* where the solid line is the boundary separating the deterministic instability (sustained-oscillatory) domain from the stable (damped-oscillatory) domain. What is most important is that the overall shape and the location of the boundary are very similar to the results from the linear transcription model in Fig. 4. The percentage of the LHS-sampled kinetic inputs generating such bifurcation diagrams is about 9 % for *HM=100 nM* as shown in Table III. But, this percentage decreases as the value of *HM* decreases; For *HM=20 nM*, this percentage is reduced to 2 %. The second transition is from the damped-oscillatory pattern to single-peaked pattern as the values of $[NF\kappa B_o]$ and *Kv* are changed from (a) to (b). In Fig. S2, we present the phase diagram for this transition.

**D. Saturating Transcription Model: Stochastic Case**

Finally, we investigated the effects of intrinsic noise on the distributions of *NF-κB* dynamic patterns in Fig. 8 and the phase diagram of Fig. 10 for the model with saturating transcription.

Using the same ensemble of input values that generated the bifurcation diagram of Fig. 10, we simulated the *NF-κB* signaling network stochastically via the Gillespie algorithm while varying the values of $[NF\kappa B_o]$ and *Kv*. The resulting *NF-κB* concentration profiles shown in Fig. 9 are very similar to those with linear transcription shown in Fig. 5. For example, at the intermediate values of $[NF\kappa B_o]$=0.04 *μM* and *Kv=4,* the deterministic network has a stable fixed point whereas the stochastic model shows noise-induced oscillations of *NF-κB,* as shown in Fig. 9(c). The power spectrum of the stochastic time-series data presented in Fig. 9(c) has a prominent peak at non-zero frequency and its *SNR* is significantly larger than one, satisfying the same criterion as before for amplified noise-induced oscillation. When the values of $[NF\kappa B_o]$ and *Kv* are set to their largest or smallest values, the intrinsic noise does not dramatically change the dynamics of *NF-κB*; a limit cycle remains unchanged in Fig. 9(a) and the noise overwhelms the signal in Fig. 9(e). Therefore, we conclude the stochastic temporal profiles and their power spectra for the saturating transcription model are very similar the linear transcription model results.

Using the criterion for noise-induced oscillations (see methods), we constructed a stochastic version of the bifurcation diagram for the saturating transcription model as shown in Fig. 10. The solid line denotes the deterministic bifurcation curve whereas the dashed line denotes the boundary line for the noise-induced oscillations. As before, this phase diagram clearly shows that noise expands the domain of oscillatory shuttling response.

Finally, we took a closer look at the effects of intrinsic noise on the distributions of *NF-κB* response patterns at the highest values of $[NF\kappa B_o]$ and *Kv*, using the same ensemble of a thousand sets of the remaining 69 input parameters. As shown in Fig. 8 and summarized in Table III, for the value of *HM* equaling100 *nM*, only about 9 % of the deterministic profiles were classified as sustained-oscillatory, whereas intrinsic noise increased this to about 30 %. This implies that phase diagrams exhibiting noise-induced oscillations, as shown in Fig. 10, can be constructed from about 30% of LHS-sampled biologically feasible input parameters. Or in terms of single-cell level experiments, about 30 % of the cells with properties sampling from the same input parameter space, could be found to exhibit oscillatory shuttling of *NF-κB*, so long as those cell's values of $[NF\kappa B_o]$ and *Kv* are in the range of intermediate to large values. We note the values of $[NF\kappa B_o]$ and *Kv* are known to be different from one cell type to another.

## III. DISCUSSION

### Experimental observations of noise-induced oscillations:

Since the modeling of the previous section predicts oscillatory shuttling of *NF-κB* across a wide range of biologically feasible kinetic parameters, a natural question is what this means for experimental observations. Our computational analysis in Fig. 10 and Table III showed only a fraction of the cells (about 30% for the model with saturating transcription with *HM=100 nM*) exhibit the noisy oscillatory shuttling of *NF-κB*. Thus, strong oscillations will likely be seen only in a fraction of cells, meaning only single-cell studies will reveal the full range of heterogeneous and stochastic response. Our computational results support the noisy oscillatory shuttling of *NF-κB* reported in the single-cell studies of Ref. [27], especially the result where 30 % of *TNF*-stimulated human cervical carcinoma cells and 70% of human S-type neuroblastoma cells demonstrated oscillations of *NF-κB* while the rest did not. Likewise, population-level studies, which only reported non-oscillatory shuttling, as found in Ref. [26, 56], could be consistent with our results. Consider averaging across millions of cells for a mixed population consisting of oscillating and non-oscillating cells and a distribution of phase delays and periods of oscillations. The resulting signal would exhibit much weaker and rapidly damped oscillations.

### Prediction of wild type behavior:

A drawback of the single-cell studies in Ref. [27] was that the *NF-κB* (RELA) and IκBα-enhanced fluorescent protein fusion construct caused *NF-κB* (RELA) and IκBα to be over-expressed relative to the normal expression level of the endogenous genes. The oscillatory shuttling observed in Ref. [27] has been criticized as abnormal behavior of *NF-κB* in mutant cells [32, 33]. Thus, an unresolved question is what the *NF-κB* shuttling dynamics would be in wild type single cells, which is a difficult experimental challenge. Our Figs. 6 & 10 demonstrate a phase boundary separating noise-induced oscillations from non-oscillations in the parameter space of $[NF\kappa B_o]$ and the ratio of cytoplasmic volume to nucleus volume (*Kv*). The value of Kv for a typical cell falls between 3 to 10. Thus, from Figs. 6 & 10, the value of $[NF\kappa B_o]$ per cell should be greater than 0.04 $\mu M$ to induce noise-induced oscillations. A typical expression level of endogenous *NF-κB* (RELA) is known to be about 60,000 molecules per mouse embryonic fibroblast cell, which is a concentration of 0.05 $\mu M$ for a typical cellular volume of 2000 $\mu m^3$. The expression level of RELA-enhanced fluorescent protein is known to be eight-fold higher than that of endogenous RELA, whose concentration is about 0.4 $\mu M$ [57]. Thus for both endogenous and fluorescent protein-enhanced expression levels of RELA, the concentrations of *NF-κB* are within the domain of noise-induced oscillatory shuttling in Figs. 6 & 10. Note that the results shown in Table III imply noise-induced oscillations will be observed in only about one third of the cells (i.e., 30 % of the LHS-sampled input parameters for the value of *HM* equaling 100 *nM*); the remaining cells would exhibit non-oscillatory responses. Thus, our computational analysis predicts that a non-negligible fraction of wild-type cells should exhibit noise-induced oscillatory shuttling patterns of *NF-κB*, if examined at the single-cell level.

**Model's applicability to various stimulus modes:**

With certain limitations, our computational analysis can be applied to the dynamic response of *NF-κB* when initiated by various stimulants such as $TNF\alpha$ and Lipopolysaccride (*LPS*). Our model is restricted to a very small autonomous signaling module, which encompasses only *IKK*, *NF-κB*, $I\kappa B\beta$, and *NF-κB*-dependent genes such as $I\kappa B\alpha$, $I\kappa B\varepsilon$ and *A20*. Thus, signal processing upstream of *IKK*, between it and the cell membrane receptors, is ignored. Fortunately, the signaling pathways of various stimulants have a common endpoint at *IKK*, which is the input to our *NF-κB* network, even though different stimulants activate different "intermediate" signaling pathways [35]. This makes our present analysis broadly applicable to the dynamic response of the *NF-κB* signaling network initiated by various stimuli, so that noise-induced oscillatory shuttling of *NF-κB* could be observed universally in response to *TNFα* or *LPS* or other kinds of stimulation. But, the kind of response we have modeled requires a constant influx of activated *IKK* protein, to continually stimulate the *NF-κB* signaling network. In reality, negative regulators may act on the "intermediate" signaling pathways between *IKK* and the cell membrane receptors to cease the inflow of activated *IKK*, likewise terminating *NF-κB* activity within the portion of the network we model [45, 58, 59]. For a particular stimulant, if a steady inflow of activated *IKK* into the *NF-κB* signaling network cannot be justified for a long duration of time, our modeling results could still be valid for whatever time scale the activated *IKK* concentration is maintained.

**Model's applicability to various cell types:**

Our computational analysis may be applicable to various cell types, with certain limitations. The computational model was originally parameterized for *NF-κB* response in *TNFα*-stimulated mouse embryonic fibroblast cells in Ref. [26]. Since this network is one of the most evolutionarily conserved biological modules and is found universally in mammals [35-37, 58], one can safely assume that both the topology of the *NF-κB* signaling network and the associated biochemical reactions are also conserved across many different species of mammals. However, one also has to keep in mind that extrinsic and intrinsic noise give rise to large phenotypic variation across a population of isogenic cells. Thus there could be large cell-to-cell variation in quantities represented as inputs to our model, e.g., from cells being in different phases of the cell-cycle and having different basal level of proteins [23, 31]. One can hypothesize that evolutionarily conserved biochemical reactions in the *NF-κB* signaling network will function in a relatively narrow range of kinetic rates across different species, and this interspecies variation in intracellular kinetic conditions can be as small as the intraspecies variation. If this hypothesis is valid, then our computational analysis should be applicable to *NF-κB* response in various mammalian cell types.

**Requirement for full understanding of the functionality of noise-induced oscillation of *NF-κB* on gene expression profiles:**

A key outstanding question is what is the biological functionality of an oscillatory shuttling response of *NF-κB?* Apart from its relation to gene expression, the functionality may be as simple as hypothesized in Ref. [31], a renewal of the *NF-κB* signaling network to freshly respond to a new signal as well as a cellular protection mechanism from high dosage-induced damage. In this regard, our computational analysis reveals that single cells could take advantage of stochastic fluctuations in intracellular biochemical reactions to enhance the probability of noise-induced oscillatory shuttling of *NF-κ*B for robustness of cellular function. However, when attempting to link *NF-κB* response to *NF-κB*-dependent gene expression profiles, the question is unresolved at present [27, 31-33]. It has been hypothesized and reported that shuttling response of *NF-κB* may be a controlling factor responsible for cell-type-specific and stimulus-specific gene expression profiles [27, 32, 33, 35, 36]. Based on non-oscillatory expression profiles of *NF-κB*-dependent genes, it has also been suggested there is no biological functionality for oscillatory shuttling of *NF-κB* [32]. But we argue that the expression profiles of *NF-κB*-dependent genes involve the dynamic interplay of multiple regulatory modules such as the *NF-κB* signaling network, the signaling network of chromatin remodeling responsible for DNA accessibility, and the signaling networks of other transcription factors required for cooperative and synergistic transcription [40-44, 50]. Our computational analysis facilitates a better understanding of the effects of noise on one of the crucial modules in this multi-regulatory machinery of gene expression. But understanding all the underlying mechanisms of *NF-κB*-dependent gene expression profiles, e.g., an expression profile of IFNβ, would require study of each of the pertinent regulatory modules, including signal-induced histone modifications and the shuttling dynamics of other transcription factors such as IRF3/7 and AP1/June, and an understanding of their dynamic interplay.

As a corollary question, what would be the functionality or potential benefit of a heterogeneous *NF-κB* response across single cells? A diverse and unharmonious response across single cells, especially in the presence of environmental stress, may seem undesirable or ineffective in fighting disease. We can suggest a possible answer to this puzzling question from microbial cell populations. Phenotypic individuality and heterogeneity observed in a bacterial population, driven by the cell cycle, cell ageing and epigenetic regulation, is thought to bolster the fitness of the population at times of stress [60-63]. For example, Ref. [60] demonstrated that a fraction of a genetically identical microbial population survive exposure to antibiotic treatment because of preexisting heterogeneity in the population, despite the persistent bacteria being sensitive to antibiotics. Thus, heterogeneous and individualistic shuttling response of *NF-κB* in single cells may grant a similar benefit of enhanced fitness to the mammalian host.

**Future directions**

With respect to the *NF-κB* signaling network, we list the following questions as outstanding issues. How do transcriptional time-delayed negative feedback loops for A20 and *IκB* affect the noise-induced oscillatory shuttling of *NF-κB* [64]? Do individual cells exhibit different *NF-κB* responses, when responding to different stimuli such as LPS or TNFα, just as reported in population level studies [45, 56] and how could that be modeled? How do we model stimulus-activated chromatin remodeling dynamics and couple it to the *NF-κB* signaling network in a unified computational framework, in order to explain *NF-κB*-dependent gene expression profiles?

## IV. MODELS & METHODS

### A. Deterministic and stochastic dynamical models of a comprehensive *NF-κB* signaling network in Figure 1

*NF-κB* is a stimulus-responsive pleiotropic gene-regulating protein [35-37]. In resting cells, *NF-κB* forms a protein complex with inhibitor proteins such as *IκBα, IκBβ,* and *IκBε,* and the *IκB-NF-κB* protein complexes are present at much higher concentration in the cytoplasm than the nucleus because the nuclear localization sequence (NLS) on a *NF-κB* protein is partially obscured by a *IκB* protein [35-37]. In cells incited by *TNFα, LPS,* or antigens, each stimulus activates their respective signaling pathway, converging to a shared endpoint, i.e., *IKK* activation. Activated *IKK* mediates the proteolysis of *IκB* protein in the *IκB-NF-κB* protein complex, resulting in the exposure of the NLS on the *NF-κB* protein and eventual migration of *NF-κB* into the nucleus. Nuclear *NF-κB* promotes the transcription of a stimulus-specific set of genes. *IκBα, IκBε* and *A20* are *NF-κB*-dependent genes and the newly synthesized *IκB* and *A20* form an auto-regulatory negative feedback loop. *IκB* binds to *NF-κB* and sequesters it in the cytoplasm whereas *A20* terminates *NF-κB* signaling by inactivating *IKK* activation [35-37, 58].

Combining all the essential components involved, we constructed a small, autonomous, yet comprehensive *NF-κB* signaling network, by combining two published computational models from Ref. [26, 39, 45]. This aggregate *NF-κB* network is presented in Fig. 1 and consists of 28 biochemical species and 70 kinetic reactions, as well as a single conserved quantity of total *NF-κB* present in the cell.

This signaling network can be readily translated into either a deterministic ODE model or a stochastic model. The functional forms of binary reactions are modeled as simple bilinear terms. The nominal values of 70 kinetic reaction parameters and the total amount of *NF-κB* are provided in Table I. Prior to stimulation, we ran the deterministic or stochastic model for about 30 hours with a single initial condition, i.e., with the total amount of *NF-κB* as the only non-zero input, equilibrating all the dynamical components in the cell. We then simulate a persistent stimulation by setting a fixed level of activated *IKK* concentration. The stochastic model also requires a value for total cell volume, which was fixed at 2000 $\mu m^3$. The standard Gillespie algorithm was used to simulate the stochastic model [52] with computational enhancements due to Ref. [65]

**B. Saturating transcription model**

The transcription process in a eukaryotic cell is very complex and typically involves multiple steps: a transcription factor binding to DNA followed by recruitment of RNA polymerase II and other factors to DNA, elongated complex formation, transcriptional pause, and mRNA splicing by splicosomes [66]. Since it seems infeasible to fully capture this process in our model, we attempt to capture the most essential with a two-step transcription model [67]:

$DNA + TF \xleftrightarrow{K_{ON}\cdot[TF];K_{OFF}} DNA^* \xrightarrow{K_{TRANS}} DNA + mRNA$ where *TF* is the transcription factor, $K_{ON}$ is an association rate of *TF* with *DNA*, $K_{OFF}$ is a dissociation rate of *TF* away from *DNA*, and *DNA** represents the activated state of a gene that produces "spliced" mRNA with a mRNA synthesis rate $K_{TRANS}$. Note that each of the reactions in the two-step model actually involves multiple hidden elementary reactions.

To further simplify the two-step transcription model to an effectively one-step process, we assume $K_{OFF}/K_{ON} >> 1$ and adiabatically eliminate the intermediate component *DNA**. The resulting one-step model is $DNA \xrightarrow{K_{TRANS}^{eff}} DNA + mRNA$ where $K_{TRANS}^{eff} = K_{TRANS}\cdot[TF]/([TF]+HM)$ and $HM = K_{OFF}/K_{ON}$. The effective mRNA synthesis rate $K_{TRANS}^{eff}$ increases proportionally to $K_{TRANS} \cdot [TF]/HM$, when $[TF] << HM$, and saturates to a constant $K_{TRANS}$, when $[TF] >> HM$.

For convenience, we substitute $K_{TRANS}$ with $K'_{TRANS} \cdot HM$ and obtain $K_{TRANS}^{eff} = K'_{TRANS} \cdot HM \cdot [TF]/([TF] + HM)$. The scaled effective mRNA synthesis rate $K_{TRANS}^{eff}$ increases linearly with $K'_{TRANS} \cdot [TF]$, independent of the values of HM, when $[TF] << HM$, and saturates to a HM-dependent value of $K'_{TRANS} \cdot HM$. As $HM \rightarrow \infty$, this linear dependence on $[TF]$ holds for all values of $[TF]$. Thus the one-step model captures both monotonically saturating (saturation to a constant) and linear transcriptional behavior (at both $[TF] << HM$ and $HM \rightarrow \infty$). Note that the large dissociation rate of $K_{OFF} / K_{ON} >> 1$ used to justify the one-step transcription implies a large value of $HM = K_{OFF} / K_{ON}$. Thus for valid use of the saturating transcription model, we should assign a reasonably large value to *HM*.

### C. Statistical ensemble of the *NF-κB* signaling network

Individual cells, even when derived from the same parental cells, are known to respond to the same dosage of the same stimulant in a heterogeneous manner, generating a diverse phenotypic variation [23, 31, 60-63]. We used a statistical ensemble approach, discussed in Ref. [51], to generate this variation by adjusting the inputs to our model of the *NF-κB* signaling network for each instance of the ensemble. This was done by a randomized Latin Hypercube sampling (LHS) methodology for the 70 kinetic rates and the total amount of *NF-κB*, where a biologically feasible range is defined for each variable, as (*0.3Xi, 1.7Xi*) where *Xi* denotes its nominal value.

### D. Latin Hypercube sampling (LHS):

LHS is a constrained Monte Carlo sampling scheme. Unconstrained Monte Carlo sampling scheme samples random points from the assumed joint probability function of the input variables. The MC samples are used to estimate the distribution of the model's response. But, for hi-dimensional spaces (many input variables), reliable MC sampling requires very large sample sizes. LHS yields accurate estimates with a smaller number of samples. Suppose that the model has K inputs and we desire N samples. LHS selects N different values for each of the K variables such that the range of each variable is divided into N non-overlapping intervals on the basis of equal probability. One value from each interval is selected at random with respect to the assumed probability density in the interval. The N values thus obtained for the first kinetic rate variable are paired in a random manner (equally likely combinations) with the N values of the second kinetic rate variable. These N pairs are combined in a random manner with the N values of the third kinetic rate variable to form N triplets, and so on, until N K-tuples are formed. These N K-tuples comprise one sample or one set of inputs for the ODE or stochastic simulation [68].

### E. Criteria for noise-induced oscillation

The power spectra we presented were calculated by taking a Fourier-transform of the stochastic time-series data of a nuclear *NF-κB* concentration profile, multiplying it with its complex conjugate, and by averaging over 10-100 time-series. The 10-100 runs were generated by Gillespie simulations with the same inputs but a different random number seed, which is used by the Gillespie solver to choose reactions in a Monte Carlo sense. The Signal to Noise Ratio (*SNR*) was then computed, which is defined as the ratio of the peak amplitude of the power spectrum to $\sqrt{[NF\kappa B_o]}$, which represents a typical stochastic fluctuation of protein concentration in the *NF-κB* signaling network [53, 54]. When the value of *SNR* is smaller than one, the signal cannot be distinguished from background stochastic fluctuations. Thus, the criteria for a response profile to exhibit amplified noise-induced oscillations were chosen as follows: (a) a non-zero resonant frequency ($\omega_0 \neq 0$) as the peak of the power spectrum and (b) a SNR greater than one.

## V. **SUPPORTING INFORMATION**

**Supporting information Fig. S1.** Another phase diagram showing both deterministic bifurcation and noise-induced oscillations of *NF-κB* for a monotonically saturating transcription model with *HM =100 nM.* This diagram was constructed with a different set of 69 input values as compared to Fig. 9, but exhibits similar domains of response. See the caption of Fig. 9 for a complete description.

**Supporting information Fig. S2.** Phase diagram exhibiting a transition between single-peaked (*SP*) and damped-oscillatory (*DAMP*) response of NF-κB for the case of a saturating transcription model (*HM =100 nM).* The values of the remaining 69 inputs were chosen such that the deterministic model produces a transition between *SP* and *DAMP* as a function of total *NF-κB* concentration ($[NF\kappa B_o]$) and volume ratio (*Kv*) as denoted in Table II. The solid blue line is the boundary between *SP* and *DAMP* for the deterministic ODE model of the network. The red-dashed line is the same boundary, but for the stochastic model with intrinsic noise. Unlike Fig. 9 and Fig. S1, the stochastic noise in this case does not induce oscillatory behavior, but transforms a fraction of *SP* responses near the blue boundary into noise-induced quasi-oscillatory responses (*NI-DAMP*), which do not meet two criteria needed to be labeled as noise-induced oscillations. I.e., *NI-DAMP* responses are characterized by the presence of a peak in their power spectrum at non-zero frequency ($\omega_o \neq 0$) but with a signal-to-noise ratio smaller than 1.

## VI. ACKNOWLEDGEMENTS

**Author contributions.** JJ conceived and designed the experiments. JJ and SJP performed the experiments. JJ analyzed the data. SJP and JLF edited the paper and provided a nurturing environment. JJ wrote the paper.

**Funding.** Sandia is a multi-program laboratory operated by Sandia Corporation, a Lockheed Martin Company, for the United States Department of Energy under contract DE-AC04-94AL85000. The authors acknowledge the Laboratory Directed Research and Development program for funding.

**Competing interests.** The authors have declared that no competing interests exist.

**FIGURE LEGENDS**

Figure 1: A two-compartment *NF-κB* signaling network. Squares stand for protein; Hexagons for mRNA; the black (solid) arrows for protein-protein interactions such as modification, degradation, and complex formation; the red and blue lines for mRNA and protein syntheses, respectively. The long dashed line separates the cell volume into two compartments, cytoplasm and nucleus. *Ai,Bi,…, Zi* represent reaction rate variables where the *i subscript* can be either α, β, or ε, which corresponds to one of three *IκB* isoforms, *IκBα, IκBβ, and IκBε*. The nominal values of the reaction rates are provided in Table I.

Figure 2: Correlation between individual scores of the four dynamic features of nuclear *NF-κB* concentration time profiles with (a) the volume ratio of the cytoplasm to the nucleus (left column), and (b) total *NF-κB* concentration (right column). A thousand time profiles of nuclear *NF-κB* concentration were generated from Latin Hypercube sampled input variables for the *NF-κB* signaling network. The red numbers are Pierson correlation coefficients. *First Max* denotes the amplitude of the first max of the time profile; *Damp* is the average slope between two adjacent oscillation peaks; *Phase* is the delay time from stimulation to first peak; Period is the average time between oscillation peaks. The lines in the graphs are only for guiding the eye.

Figure 3: Distributions of four dynamic *NF-κB* shuttling patterns for sets of input variables which were (a) all LHS-sampled, and (b) LHS-sampled except for total *NF-κB* concentration ($[NF\kappa B_o]$) and volume ratio of cytoplasm to nucleus (*Kv*) being set to large values. This is for a linear transcription model (*HM*=∞) and deterministic (ODE) simulations of the network of Fig. 1. The yellow area represents sustained-oscillation; red area is damped-oscillation; black area is monotonically saturating response; blue area is single-peaked response. For (a), $[NF\kappa B_o]$ and *Kv* are LHS-sampled from the intervals (*0.01 μM, 0.1 μM*) and (1, 10), respectively. For (b), $[NF\kappa B_o]=0.1\ \mu M$ & $Kv=10$. Each column of shuttling patterns was derived from a thousand simulations of sampled sets of the 71 input variables.

Figure 4: Bifurcation phase diagram of the shuttling response of *NF-κB* as a function of two key parameters: total NF-κB concentration ($[NF\kappa B_o]$) and volume ratio of cytoplasm to nucleus (*Kv*). This is for a linear transcription model (*HM*=∞) and deterministic (ODE) simulations of the network. NF-κB shuttling dynamics in the *OSC* domain bounded by the blue line is unstable, i.e., sustained-oscillations result. In the *DAMP* domain, *NF-κB* response is stable, i.e., damped oscillations result. In the inset (a), an unstable time profile of NF-κB is shown for $[NF\kappa B_o]=0.1$ *μM* & *Kv=9* (square). In the inset (b), a stable time profile of *NF-κB* is shown for $[NF\kappa B_o]=0.06$ *μM* & *Kv=4* (circle). The remaining 69 kinetic parameter values were chosen from one of the thousand input sets in Fig. 3(b) that generated sustained-oscillatory shuttling of NF-κB at the largest values of $[NF\kappa B_o]=0.1$ *μM* & *Kv=10.*

Figure 5: Effects of intrinsic noise on the temporal profiles of nuclear *NF-κB* concentration and their power spectra for the case of a linear transcription model (*HM*=∞*)*. The graphs shown in (a), (c), and (e) are temporal profiles of nuclear NF-κB concentration using a deterministic model (thin blue lines) and a stochastic model (thick red lines) of the network in Fig. 1. The graphs in (b), (d), and (f) present the power spectra obtained from a Fourier-Transform of the stochastic time-series data in (a), (c), and (e). The values of $[NF\kappa B_o]$ and *Kv* are $[NF\kappa B_o]=0.1$ *μM* & *Kv=9* (squares in Fig. 4) for (a) and (b), $[NF\kappa B_o]=0.06$ *μM* & *Kv=4* (circles in Fig. 4) for (c) & (d), and $[NF\kappa B_o]=0.04$ *μM* & *Kv=2* for (e) & (f). The signal-to-noise ratio (*SNR*) is the ratio of the peak amplitude of a power spectrum to $\sqrt{[NF\kappa B_o]}$. The values of the remaining 69 input values are the same as in Fig. 4, as listed in Table I.

Figure 6: Phase diagram exhibiting both a deterministic bifurcation and noise-induced oscillatory shuttling of *NF-κB* based on the same two key parameters as before: total *NF-κB* concentration and volume ratio of cytoplasm to nucleus. This is for a linear transcription model (*HM*=∞). *DAMP* and *OSC* denote domains in which the deterministic ODE model yields damped oscillatory and sustained oscillatory *NF-κB* responses, respectively. In the noise-induced oscillations (*NIO)* domain, the stochastic model exhibits noisy quasi-periodic shuttling of *NF-κB* for input values where the deterministic dynamics would be stable. The *NIO* are characterized by two criteria: a signal-to-noise ratio of their power spectrum greater than one and a non-zero frequency of their power spectrum. The red dashed line separates the *NIO* domain from the non-oscillatory domain (*DAMP*) while the blue line divides the deterministic instability domain (*OSC*) from the stochastic noise-induced oscillatory domain (*NIO)*. The remaining 69 input values are the same as in Figs. 4 and 5 and Table I.

Figure 7: Illustration of long-excursion deterministic phase trajectories in the parameter space of the nuclear NF-κB concentration and cytoplasmic IκBα concentration for the case of a linear transcription model (*HM*=∞). In (a) is a phase trajectory of a limit cycle which flows along the direction of the arrow for $[NF\kappa B_o]=0.1\ \mu M$ (total concentration) & $Kv=9$ (squares in Figs. 4 & 5). In (b) is a phase trajectory of stable dynamics for $[NF\kappa B_o]=0.06\ \mu M$ & $Kv=4$ (circles in Figs. 4 & 5). In (c) is a phase trajectory of stable dynamics for $[NF\kappa B_o]=0.04\ \mu M$ & $Kv=2$ (triangles in Fig. 5). The remaining 69 input values are the same in Figs. 4, 5, and 6 and Table I.

Figure 8: Distributions of the shuttling response of *NF-κB* for a saturating transcription model; contrast to Fig. 3. The three bars on the left are for LHS-sampling of all inputs while the three bars on the right are for LHS-sampling of all inputs except total *NF-κB* concentration ($[NF\kappa B_o]$) and volume ratio of cytoplasm to nucleus (*Kv*) being set to large values. *HM* is defined as the ratio of the dissociation rate of *NF-κB* from the *DNA* to its association rate with the *DNA. The* values *HM = 20 nM, 100 nM,* and 1000 *nM* set the threshold concentration of *NF-κB* for transcriptional saturation. The yellow area represents sustained-oscillations; red area is damped-oscillations; blue area is single-peaked response. The monotonically saturating response, which was present for the linear transcriptional model, is absent from these distributions. For (a), $[NF\kappa B_o]$ and *Kv* are LHS-sampled from the intervals (*0.01 μM, 0.1 μM*) and (1, 10), respectively. For (b), the values of $[NF\kappa B_o]$ and *Kv* are fixed at the largest values in the intervals: $[NF\kappa B_o]=0.1\ \mu M\ \&\ Kv=10$. As in Fig. 3, each column of shuttling patterns was derived from a thousand simulations of sampled sets of the 71 input variables.

Figure 9: Effects of intrinsic noise on the temporal profiles of nuclear *NF-κB* concentration and their power spectra for a monotonically saturating transcription model (*HM=0.1 μM);* contrast to *Fig. 5.* The graphs shown in (a), (c), and (e) depict temporal profiles of nuclear NF-κB concentration from the deterministic ODE model (thin blue lines) and from the stochastic model (thick red lines). The graphs in (b), (d), and (f) are power spectra obtained from the Fourier-Transform of the stochastic time-series data in (a), (c), and (e). The values of $[NF\kappa B_o]$ and $Kv$ are $[NF\kappa B_o]=0.1\ \mu M$ & $Kv=10$ (a square in Fig. 10) for (a) and (b), $[NF\kappa B_o]=0.04\ \mu M$ & $Kv=4$ (a circle in Fig. 10) for (c) & (d), and $[NF\kappa B_o]=0.02\ \mu M$ & $Kv=2$ (a triangle in Fig. 10) for (e) & (f). The signal-to-noise ratio (*SNR*) is the ratio of the peak amplitude of a power spectrum to $\sqrt{[NF\kappa B_o]}$ . The values of the remaining 69 inputs are listed in Table I.

Figure 10: Phase diagram exhibiting both a deterministic bifurcation and noise-induced oscillatory shuttling of *NF-κB* for a monotonically saturating transcription model (*HM=100 nM)*; contrast to Fig. 6. The input parameter values are the same as in Fig. 9. The figure annotations are the same as described in Fig. 6.

Table I. Biochemical reactions & associated reaction rates in our computational model of the NF-κB signaling network. The reaction rates labeled with [1] are from Ref. [39], those labeled [2] are from Ref. [45], those labeled [3] use an average value between those in Ref. [39] & Ref. [45]. Column *I* is the kinetic parameter, *II* is its units, *III* is its nominal value, and *IV* is the reference. *HM=∞* indicates the set of values used for Figs. 4, 5, 6, and 7 for the linear transcription model. *HM=0.1 μM* denotes the set of values used for Figs. 9 and 10 for a monotonically saturating transcription model. The units for [a] are $\mu M^{-1}s^{-1}$, for [b] are $s^{-1}$, for [c] are $\mu M\ s^{-1}$, and for [d] are μM.

| Reactions | I | II | III | IV | *HM=∞* | *HM=0.1 μM* |
|---|---|---|---|---|---|---|
| IKKa + IkBa → IKKa_IkBα | Aα | [a] | 0.2 | [1] | 0.1813 | 0.2243 |
| IKKa + IkBb → IKKa_IkBβ | Aβ | [a] | 0.05 | [3] | 0.02997 | 0.04994 |
| IKKa + IkBe → IKKa_IkBε | Aε | [a] | 0.05 | [3] | 0.04244 | 0.06604 |
| IKKa+IkBα_NFkB → IKKa_IkBα_NfkB | Bα | [a] | 1 | [1] | 1.024 | 1.421 |
| IKKa+IkBβ_NFkB → IKKa_IkBβ_NFkB | Bβ | [a] | 0.25 | [3] | 0.3683 | 0.2427 |
| IKKa+IkBε_NFkB → IKKa_IkBε_NFkB | Bε | [a] | 0.25 | [3] | 0.42 | 0.1575 |
| NFkBn → NFkBn + A20t | C1 | [b] | 0.0000005 | [1] | 0.000000506 | 0.000000769 |
| 0 → A20t | C2 | [c] | 0 | [1] | 0 | 0 |
| A20t → 0 | C3 | [b] | 0.0004 | [1] | 0.0002438 | 0.0004879 |
| A20t → A20t + A20 | C4 | [b] | 0.5 | [1] | 0.5807 | 0.6816 |
| A20 → 0 | C5 | [b] | 0.0003 | [1] | 0.0003769 | 0.0003 |
| IKKa_IkBα → IKKa + IkBα | Dα | [b] | 0.00125 | [2] | 0.002046 | 0.00066 |
| IKKa_IkBβ → IKKa + IkBβ | Dβ | [b] | 0.00175 | [2] | 0.0005609 | 0.002409 |
| IKKa_IkBε → IKKa + IkBε | Dε | [b] | 0.00175 | [2] | 0.002142 | 0.0009395 |
| IKKa_IkBα_NFkB → IKKa + IkBα_NFkB | Dα | [b] | 0.00125 | [2] | 0.002046 | 0.00066 |
| IKKa_IkBβ_NFkB → IKKa + IkBβ_NFkB | Dβ | [b] | 0.00175 | [2] | 0.000561 | 0.002409 |
| IKKa_IkBε_NFkB → IKKa + IkBε_NFkB | Dε | [b] | 0.00175 | [2] | 0.002142 | 0.0009395 |
| IKKa_IkBα_NFkB → IKKa_IkBα + NFkB | Eα | [b] | 0.000001 | [2] | 0.00000144 | 0.000000529 |
| IKKa_IkBβ_NFkB → IKKa_IkBβ + NFkB | Eβ | [b] | 0.000001 | [2] | 0.00000124 | 0.00000149 |
| IKKa_IkBε_NFkB → IKKa_IkBε + NFkB | Eε | [b] | 0.000001 | [2] | 0.00000064 | 0.00000138 |
| IKKa_IkBα + NFkB → IKKa_IkBα_NFkB | Fα | [a] | 0.5 | [2] | 0.3789 | 0.1593 |
| IKKa_IkBβ + NFkB → IKKa_IkBβ_NFkB | Fβ | [a] | 0.5 | [2] | 0.2135 | 0.2394 |
| IKKa_IkBε + NFkB → IKKa_IkBε_NFkB | Fε | [a] | 0.5 | [2] | 0.3528 | 0.4183 |
| IkBα_NFkB → NFkB + IkBα | Gα | [b] | 0.000001 | [2] | 0.00000064 | 0.00000105 |
| IkBβ_NFkB → NFkB + IkBβ | Gβ | [b] | 0.000001 | [2] | 0.00000044 | 0.00000113 |
| IkBε_NFkB → NFkB + IkBε | Gε | [b] | 0.000001 | [2] | 0.00000069 | 0.00000084 |
| IkBαn_NFkBn → NFkBn + IkBαn | Gα | [b] | 0.000001 | [2] | 0.00000064 | 0.00000105 |
| IkBβn_NFkBn → NFkBn + IkBβn | Gβ | [b] | 0.000001 | [2] | 0.00000044 | 0.00000113 |
| IkBεn_NFkBn → NFkBn + IkBεn | Gε | [b] | 0.000001 | [2] | 0.00000069 | 0.00000084 |
| IkBα + NFkB → IkBα_NFkB | Hα | [a] | 0.5 | [2] | 0.4593 | 0.3691 |
| IkBβ + NFkB → IkBβ_NFkB | Hβ | [a] | 0.5 | [2] | 0.7753 | 0.166 |
| IkBε + NFkB → IkBε_NFkB | Hε | [a] | 0.5 | [2] | 0.2895 | 0.6864 |
| IkBαn + NFkBn → IkBαn_NFkBn | Hα | [a] | 0.5 | [2] | 0.4593 | 0.3691 |
| IkBβn + NFkBn → IkBβn_NFkBn | Hβ | [a] | 0.5 | [2] | 0.7753 | 0.166 |
| IkBεn + NFkBn → IkBεn_NFkBn | Hε | [a] | 0.5 | [2] | 0.2895 | 0.6864 |
| NFkB → NFkBn | I1 | [b] | 0.0025 | [1] | 0.003037 | 0.002509 |
| NFkBn → NFkB | K01 | [b] | 0.00005 | [3] | 0.00005537 | 0.00004072 |
| IKKn → IKKa | K1 | [b] | 0.0025 | [1] | 0.003273 | 0.001616 |
| A20 +IKKa → A20 + IKKi | K2 | [a] | 0.1 | [1] | 0.07075 | 0.1698 |
| IKKa → IKKi | K3 | [b] | 0.0015 | [1] | 0.00202 | 0.00205 |
| 0 → IKKn | Kprod | [c] | 0.000025 | [1] | 0.000009752 | 0.00002428 |
| IKKn, IKKa, or IKKi → 0 | Kdeg | [b] | 0.000125 | [1] | 0.0001561 | 0.00006858 |
| Volume ratio of cytoplasm to nucleus | Kv | 1 | 5 | [1] | Variable | Variable |
| IkBαn_NFkBn → IkBα_NFkB | Lα | [b] | 0.01 | [1] | 0.013979 | 0.007196 |
| IkBβn_NFkBn → IkBβ_NFkB | Lβ | [b] | 0.005 | [3] | 0.001567 | 0.007097 |
| IkBεn_NFkBn → IkBε_NFkB | Lε | [b] | 0.005 | [3] | 0.006583 | 0.007048 |
| IkBα_NFkB → NFkB | Mα | [b] | 0.000025 | [1] | 0.00002837 | 0.000022 |
| IkBβ_NFkB → NFkB | Mβ | [b] | 0.000025 | [3] | 0.00003609 | 0.0000175 |
| IkBε_NFkB → NFkB | Mε | [b] | 0.000025 | [3] | 0.00000866 | 0.0000219 |
| IKKa_IkBα_NFkB → IKKa + NFkB | Pα | [b] | 0.1 | [1] | 0.12928 | 0.1486 |
| IKKa_IkBβ_NFkB → IKKa + NFkB | Pβ | [b] | 0.05 | [3] | 0.06454 | 0.01655 |
| IKKa_IkBε_NFkB → IKKa + NFkB | Pε | [b] | 0.05 | [3] | 0.08434 | 0.04706 |
| IkBαn → IkBα | Qα | [b] | 0.0005 | [1] | 0.0005123 | 0.0001987 |
| IkBβn → IkBβ | Qβ | [b] | 0.0005 | [3] | 0.0007398 | 0.0002571 |
| IkBεn → IkBε | Qε | [b] | 0.0005 | [3] | 0.0002184 | 0.0005221 |
| IKKa_IkBα → IKKa | Rα | [b] | 0.1 | [1] | 0.123 | 0.1088 |
| IKKa_IkBβ → IKKa | Rβ | [b] | 0.1 | [3] | 0.03837 | 0.07846 |
| IKKa_IkBε → IKKa | Rε | [b] | 0.1 | [3] | 0.1571 | 0.1119 |
| IkBαn_NFkBn → NFkBn | Sα | [b] | 0.000001 | [2] | 0.00000037 | 0.00000163 |
| IkBβn_NFkBn → NFkBn | Sβ | [b] | 0.000001 | [2] | 0.000001131 | 0.000001103 |
| IkBεn_NFkBn → NFkBn | Sε | [b] | 0.000001 | [2] | 0.000001037 | 0.000000358 |
| NFkBn → NFkBn + IkBαt | Uα | [b] | 0.0000005 | [1] | 0.000000279 | 0.000000333 |
| NFkBn → NFkBn + IkBβt | Uβ | [b] | 0 | [2] | 0 | 0 |
| NFkBn → NFkBn + IkBεt | Uε | [b] | 0.00000005 | [3] | 0.000000059 | 0.000000028 |
| IkBα → IkBαn | Vα | [b] | 0.001 | [1] | 0.0009786 | 0.001616 |
| IkBβ → IkBβn | Vβ | [b] | 0.001 | [3] | 0.0004871 | 0.001074 |
| IkBε → IkBεn | Vε | [b] | 0.001 | [3] | 0.00147 | 0.00113 |
| IkBα, IkBαn → 0 | Wα | [b] | 0.0001 | [1] | 0.000132 | 0.0000601 |
| IkBβ, IkBβn → 0 | Wβ | [b] | 0.0001 | [3] | 0.000133 | 0.0000826 |
| IkBε, IkBεn → 0 | Wε | [b] | 0.0001 | [3] | 0.000042 | 0.0001 |
| IkBαt → IkBαt + IkBα | Xα | [b] | 0.5 | [1] | 0.4552 | 0.4536 |
| IkBβt → IkBαt + IkBβ | Xβ | [b] | 0.5 | [3] | 0.3828 | 0.3868 |
| IkBεt → IkBαt + IkBε | Xε | [b] | 0.5 | [3] | 0.3304 | 0.4332 |
| 0 → IkBαt | Yα | [c] | 0.00000005 | [3] | 0.000000084 | 0.000000058 |
| 0 → IkBβt | Yβ | [c] | 0.000000005 | [3] | 0.00000000414 | 0.0000000077 |
| 0 → IkBεt | Yε | [c] | 0.000000005 | [3] | 0.00000000508 | 0.0000000048 |
| IkBαt → 0 | Zα | [b] | 0.0004 | [1] | 0.0003375 | 0.0005607 |
| IkBβt → 0 | Zβ | [b] | 0.0004 | [3] | 0.0002031 | 0.0004477 |
| IkBεt → 0 | Zε | [b] | 0.0004 | [3] | 0.0004742 | 0.0001471 |
| Total NFkB amount | $NF\kappa B_o$ | [d] | 0.06 | [1] | Variable | Variable |

Table II: All possible transitions between different *NF-κB* shuttling responses as a function of total *NF-κB* concentration ($[NF\kappa B_o]$) and the volume ratio of cytoplasm to nucleus (*Kv*) for the cases of both linear and monotonically saturating transcription models. For data set [A], we LHS-sampled $[NF\kappa B_o]$ and *Kv* from the intervals $[NF\kappa B_o]\in(0.01$ μM, 0.1 μM) and $Kv\in(1,10)$ as well as all the other 69 input variables and constructed distributions of the NF-κB shuttling responses for different transcriptional saturations defined by the value of *HM*, as shown in Figs. 3 & 8. For data set [B], holding the remaining 69 input values as same as the data set [A], $[NF\kappa B_o]$ and *Kv* were set to the largest values in their respective intervals, i.e., $[NF\kappa B_o]=0.1\ \mu M$ and $Kv=10$, and the distributions of the NF-κB shuttling responses were again constructed. Comparing time profiles in [A] to [B] yields a list of all transitions between individual patterns. *DAMP* stands for damped-oscillatory nuclear *NF-κB* time profiles; *OSC* for sustained-oscillatory; *SP* for single-peaked; *MS* for monotonically saturating; *HM* is the ratio of the dissociation rate to the association rate between DNA and NF-κB in the monotonically saturating transcription model

| Transitions from [A] to [B] | *HM=∞* | *HM=1 μM* | *HM=100 nM* | *HM=20 nM* |
|---|---|---|---|---|
| *DAMP → OSC* | 18 % | 11 % | 6 % | 1 % |
| *DAMP → SP* | 0 % | 0 % | 9 % | 22 % |
| *SP → DAMP* | 7 % | 1 % | 1 % | 1 % |
| *MS → DAMP* | 13 % | 0 % | 0 % | 0 % |
| *DAMP → DAMP* | 60 % | 82 % | 68 % | 23 % |
| *SP → SP* | 0 % | 0 % | 13 % | 0 % |
| *OSC → OSC* | 2 % | 6 % | 3 % | 1 % |

Table III: Noise-enhanced probability of seeing an oscillatory shuttling response of *NF-κB* at the largest values of $[NF\kappa B_o]=0.1\ \mu M$ & $Kv=10$. For each combination of transcription model and dynamical model (ODE and stochastic), we LHS-sampled a hundred different configurations of the remaining 69 input values and computed the percentage of those which exhibited oscillatory shuttling. *NIO* stand for noise-induced oscillations. Inclusion of intrinsic stochastic noise in the model clearly enhances the likelihood of an oscillatory response.

| Transcription model | Percentage of *OSC* from the deterministic model | Percentage of *NIO* from the Stochastic model |
|---|---|---|
| Linear transcription ($HM=\propto$) | 20 % | 99 % |
| Monotonically saturating transcription ($HM=100\ nM$) | 9 % | 31 % |

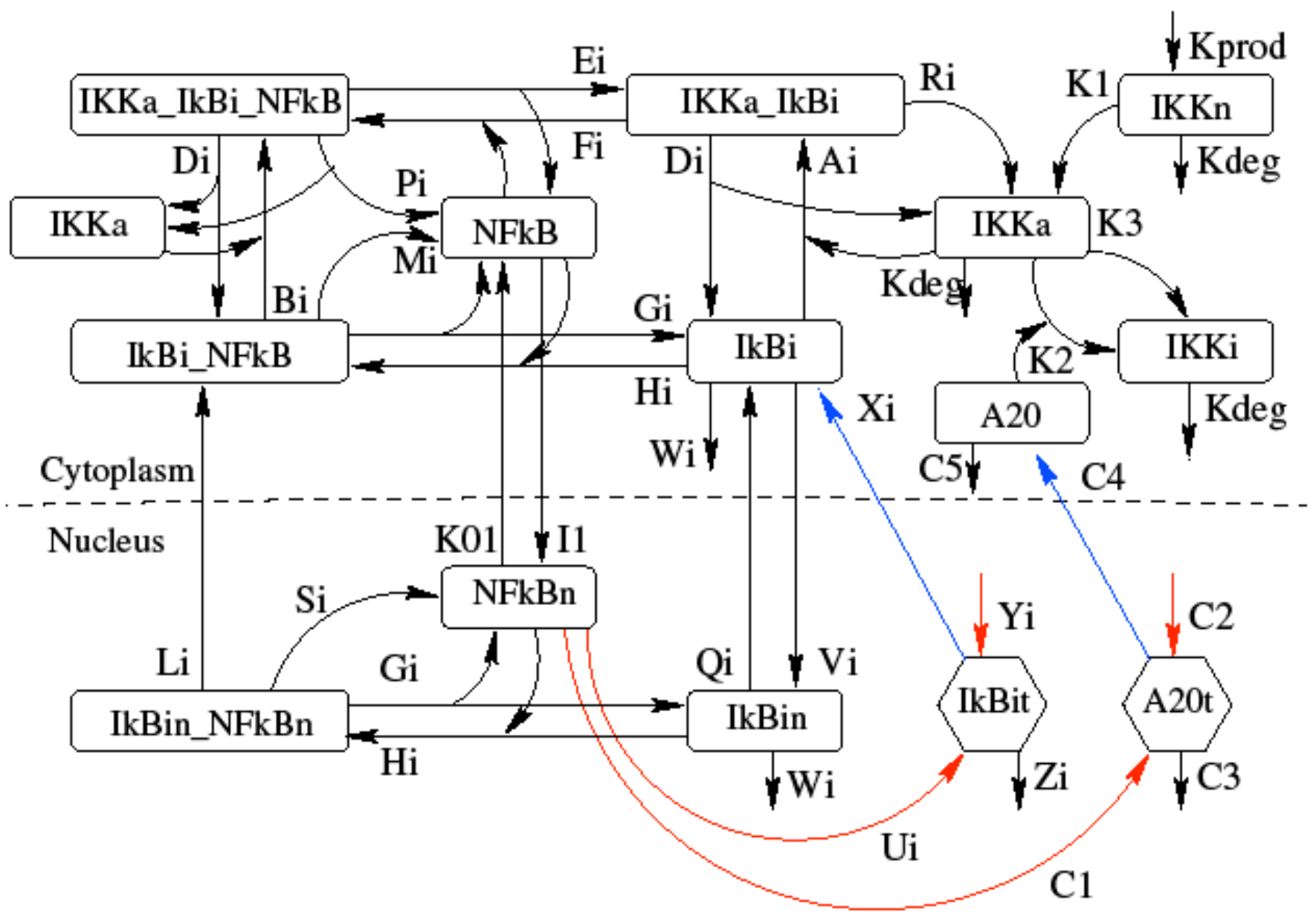

Figure 1

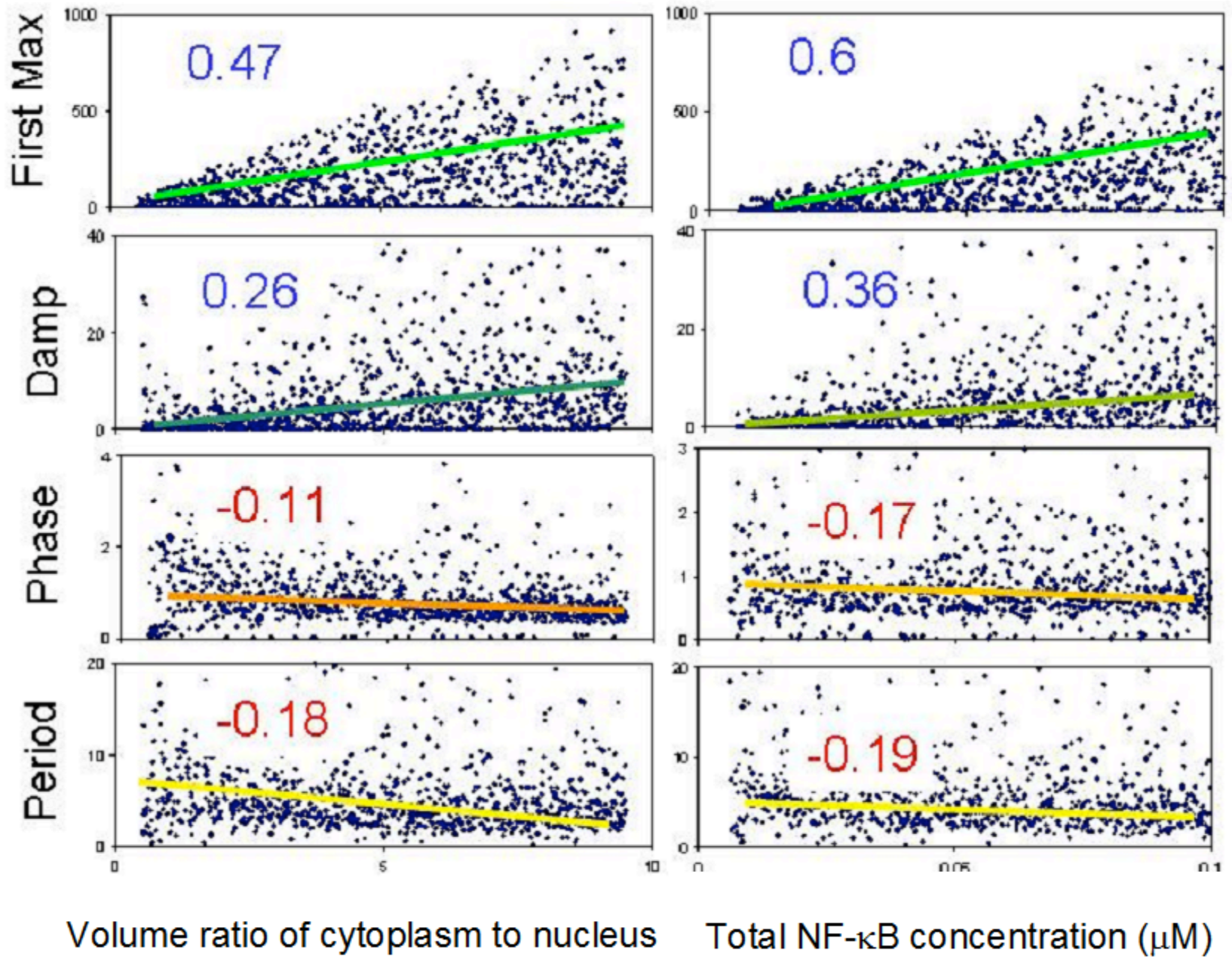

Figure 2

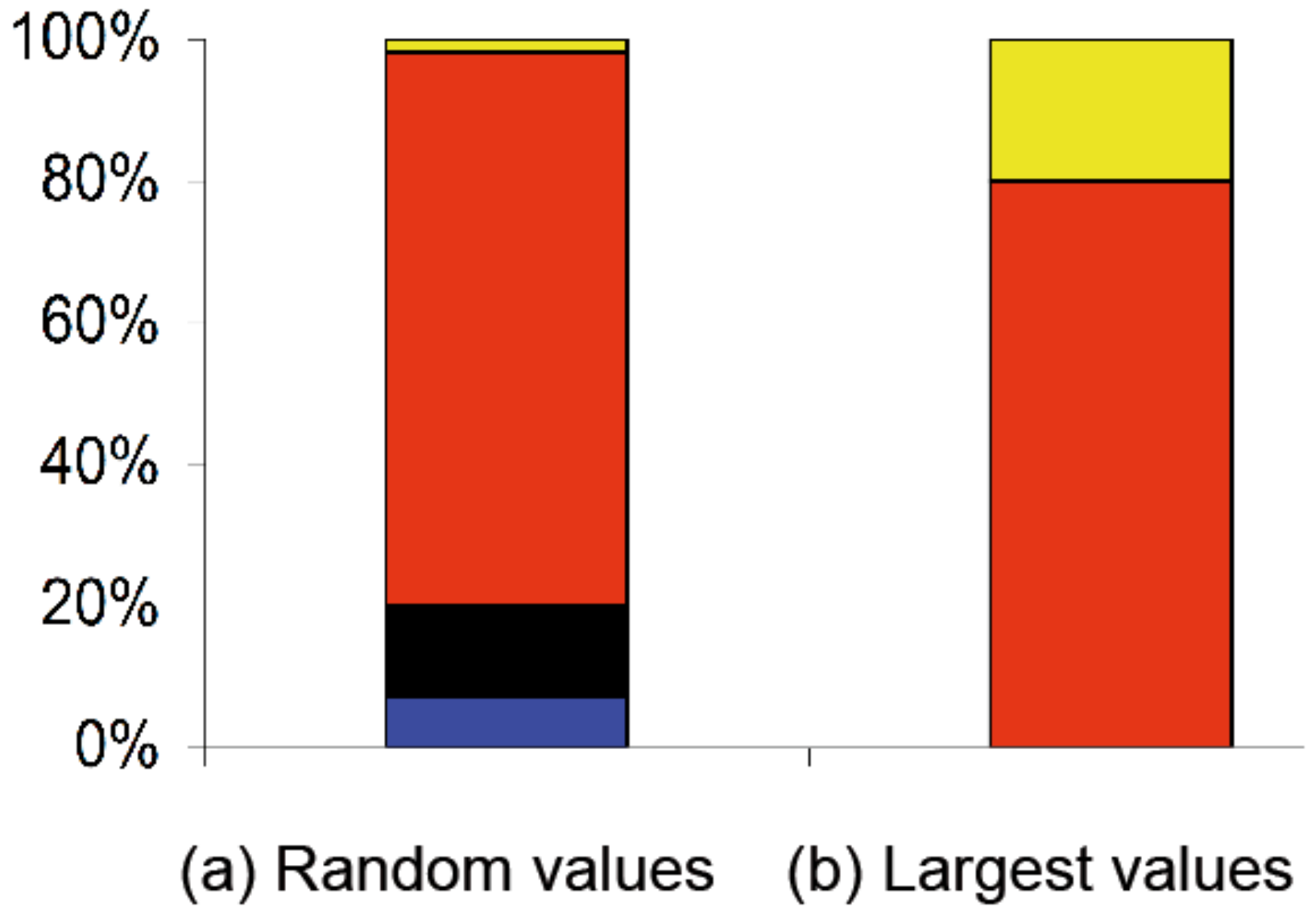

Figure 3

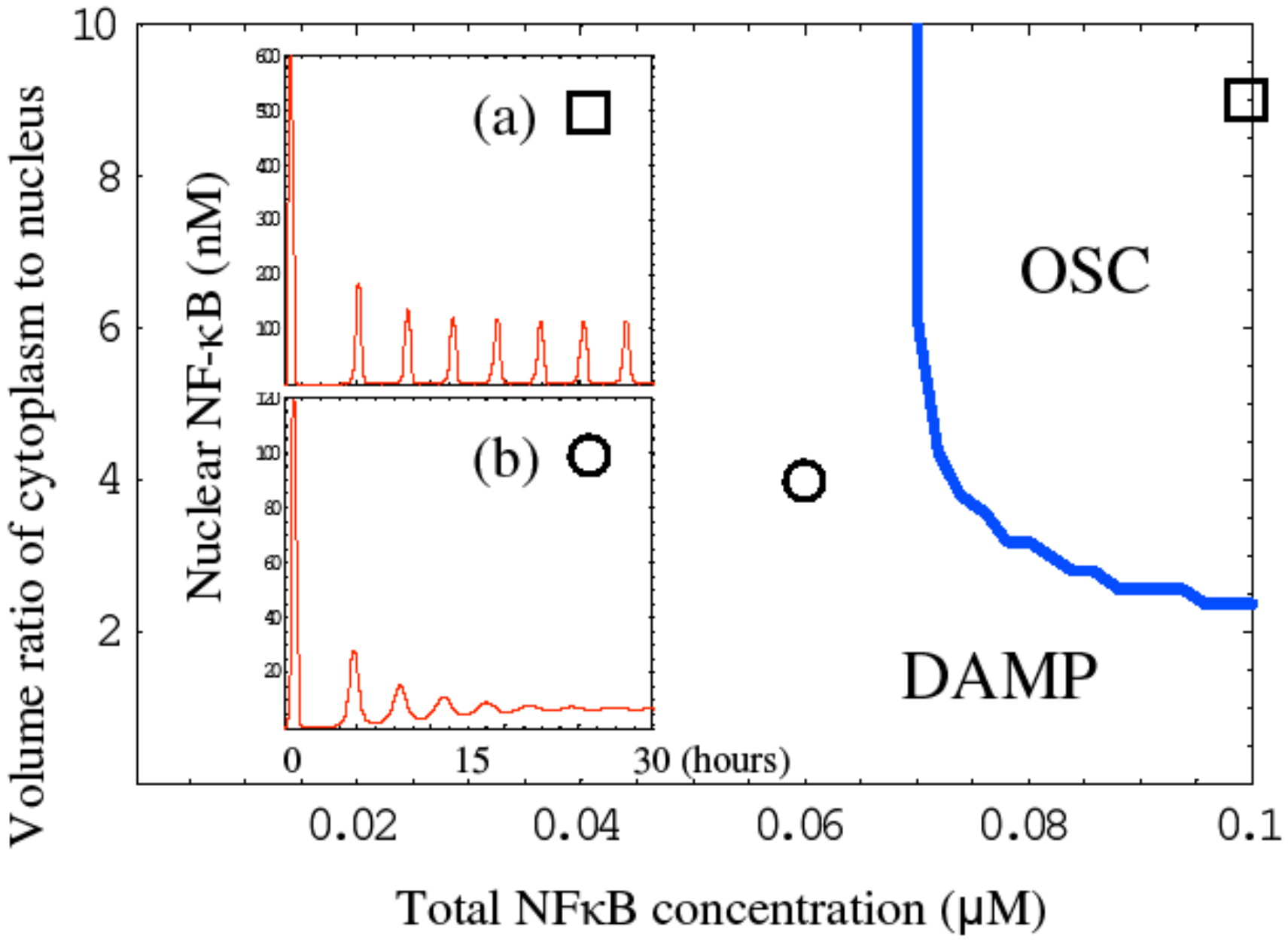

Figure 4

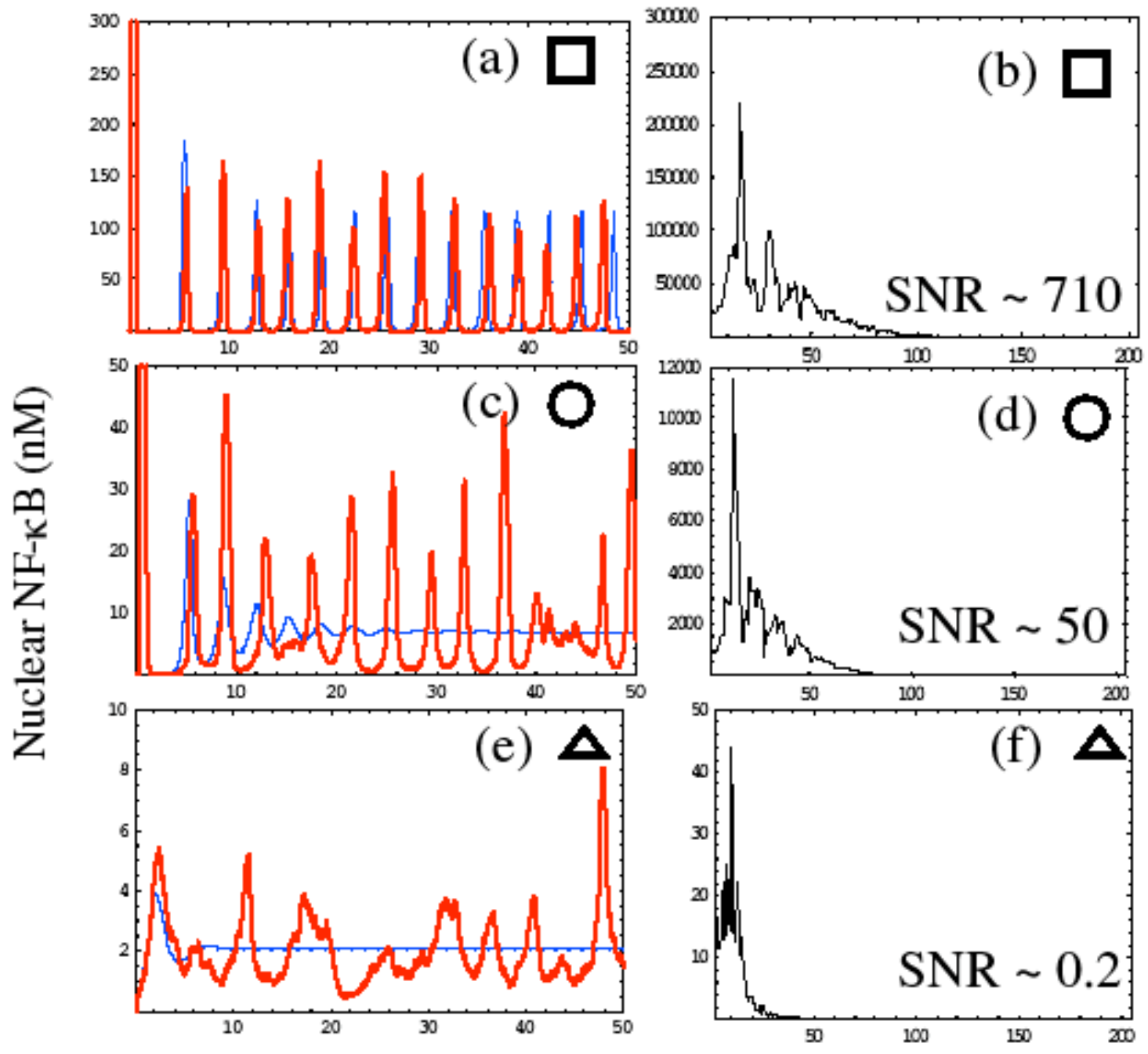

Figure 5

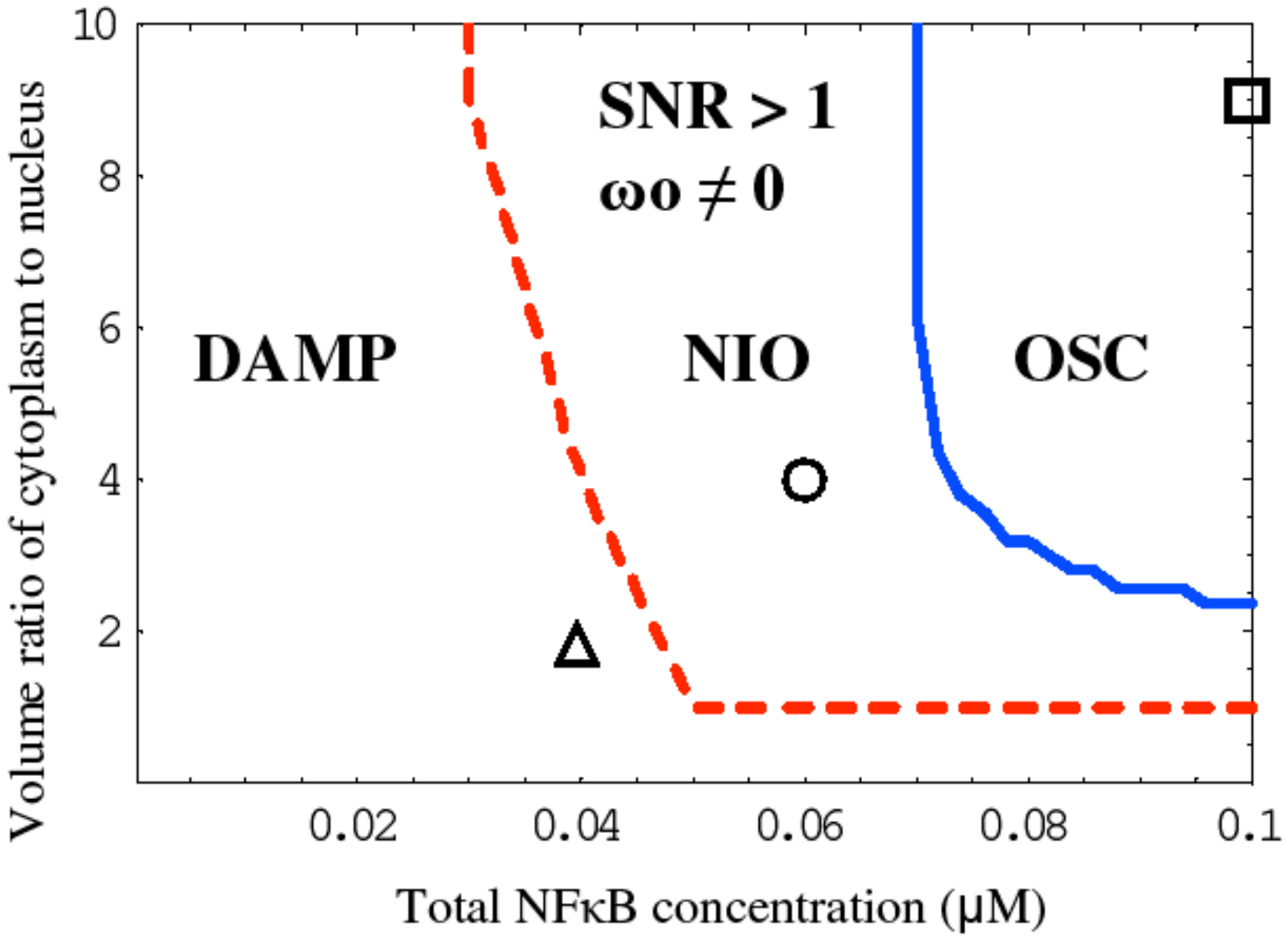

Figure 6

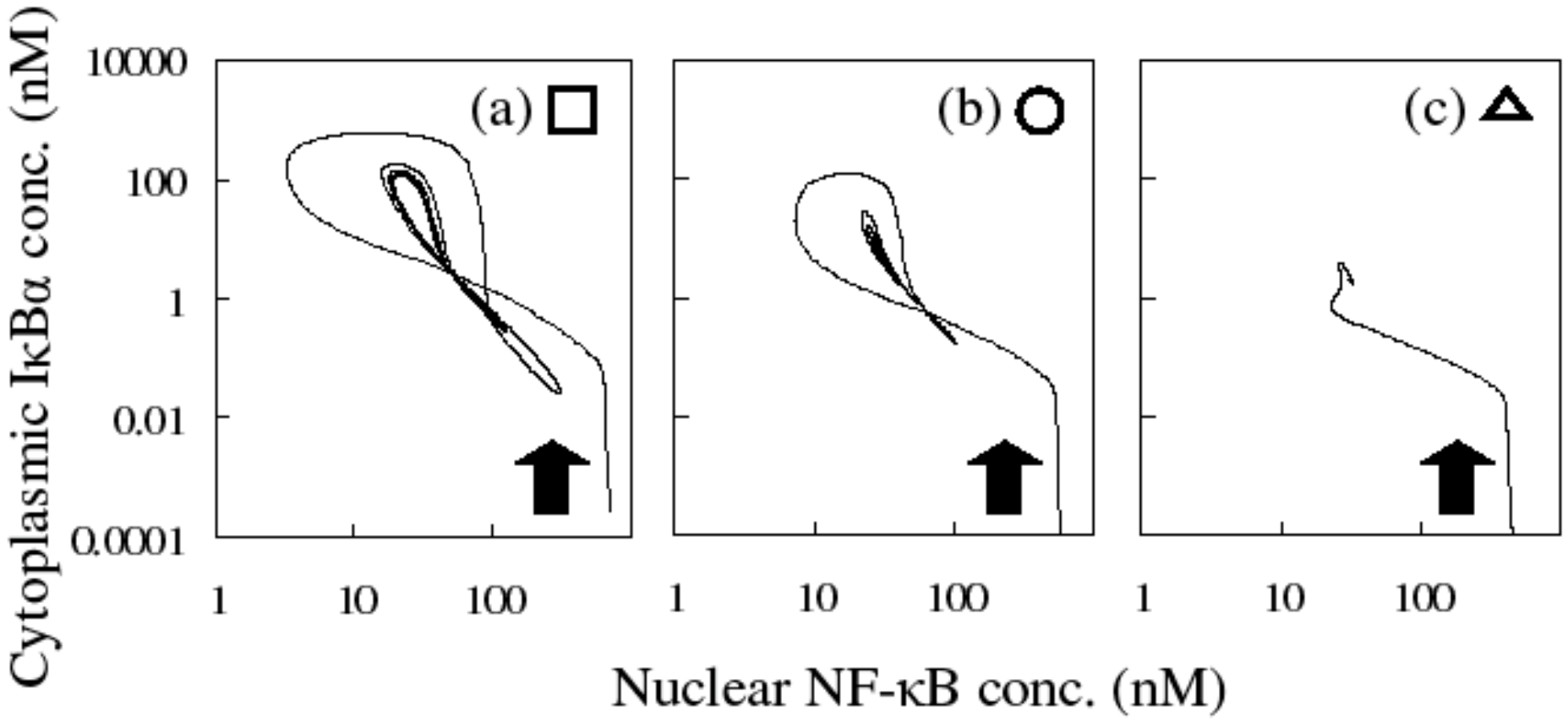

Figure 7

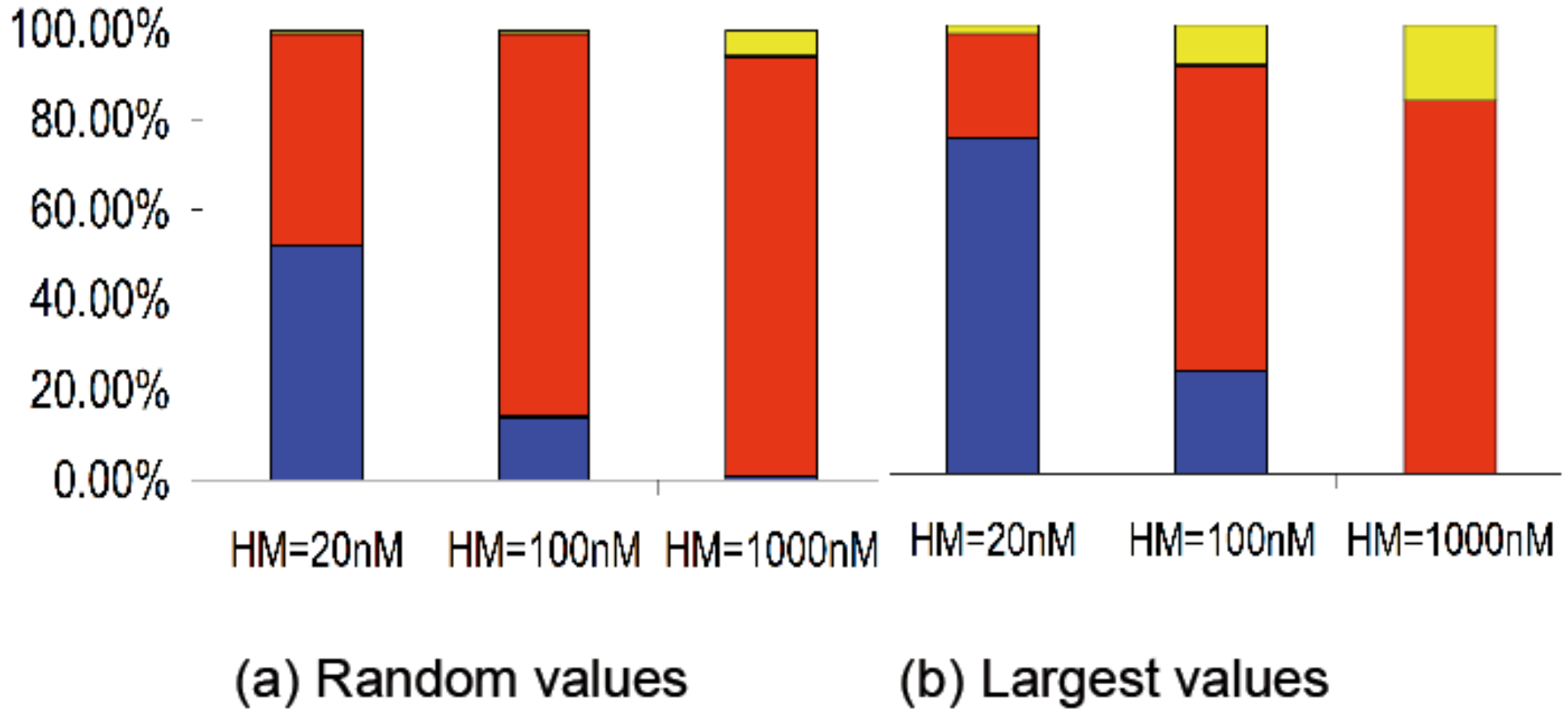

Figure 8

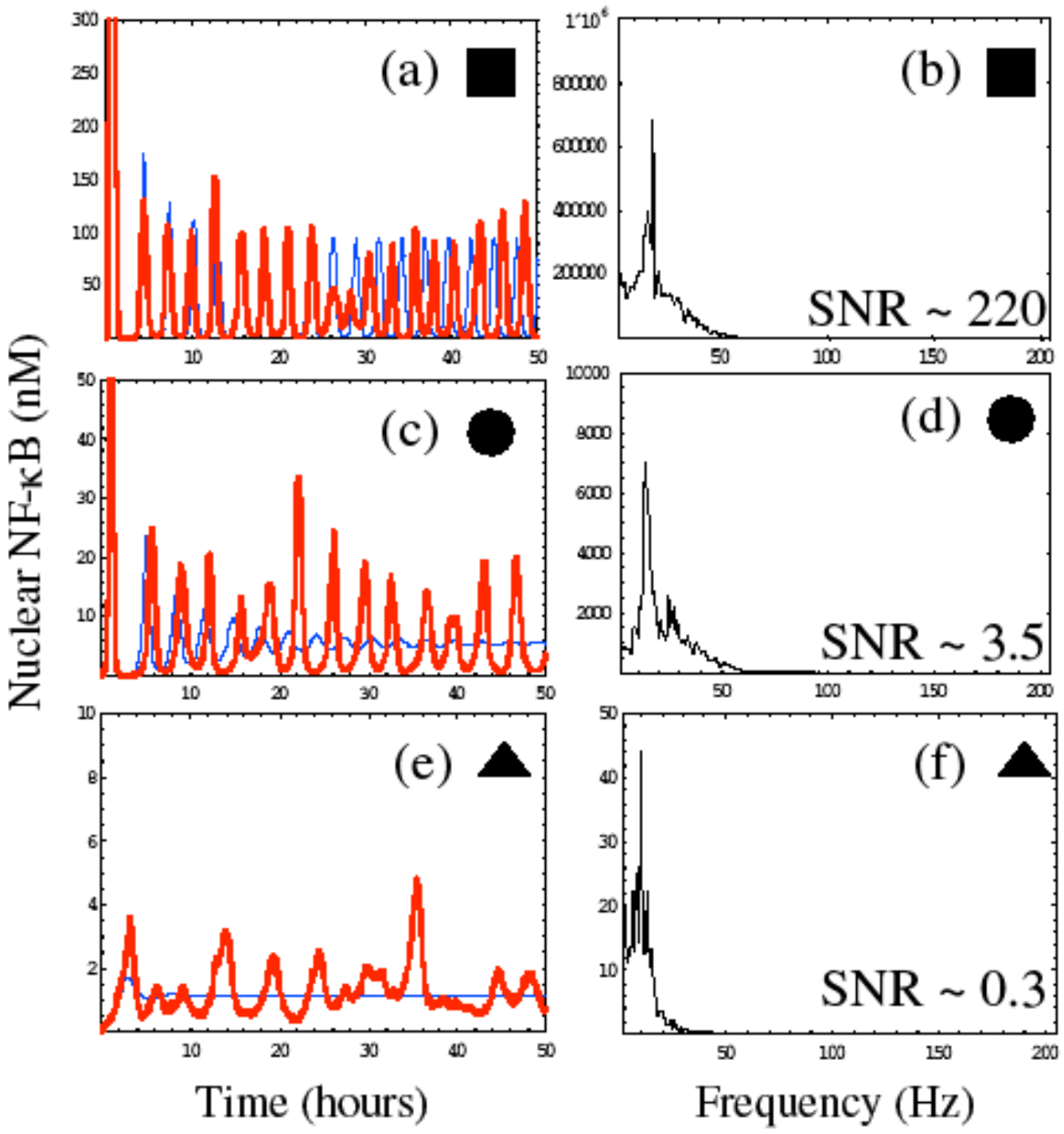

Figure 9

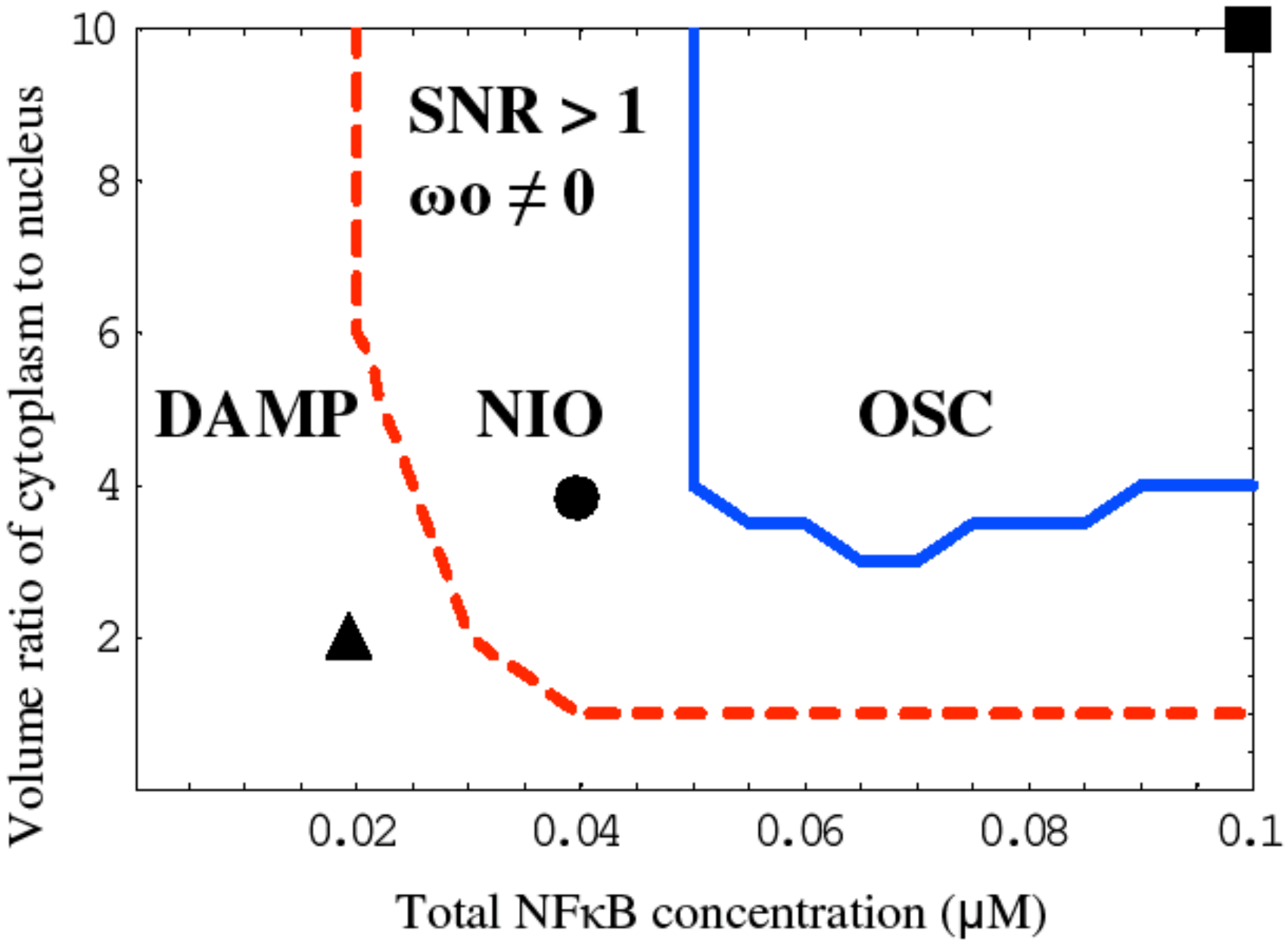

Figure 10

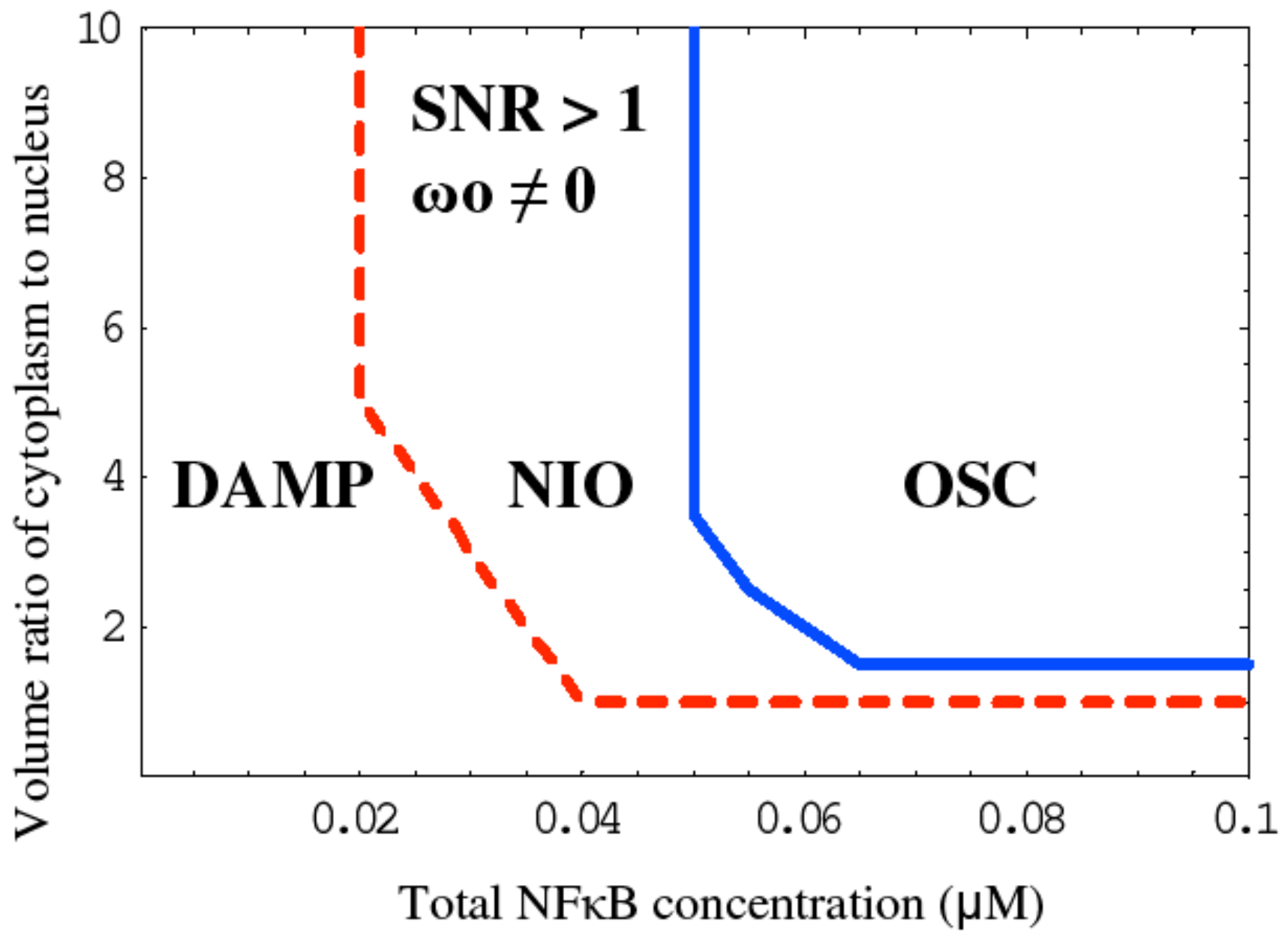

Supporting information figure S1

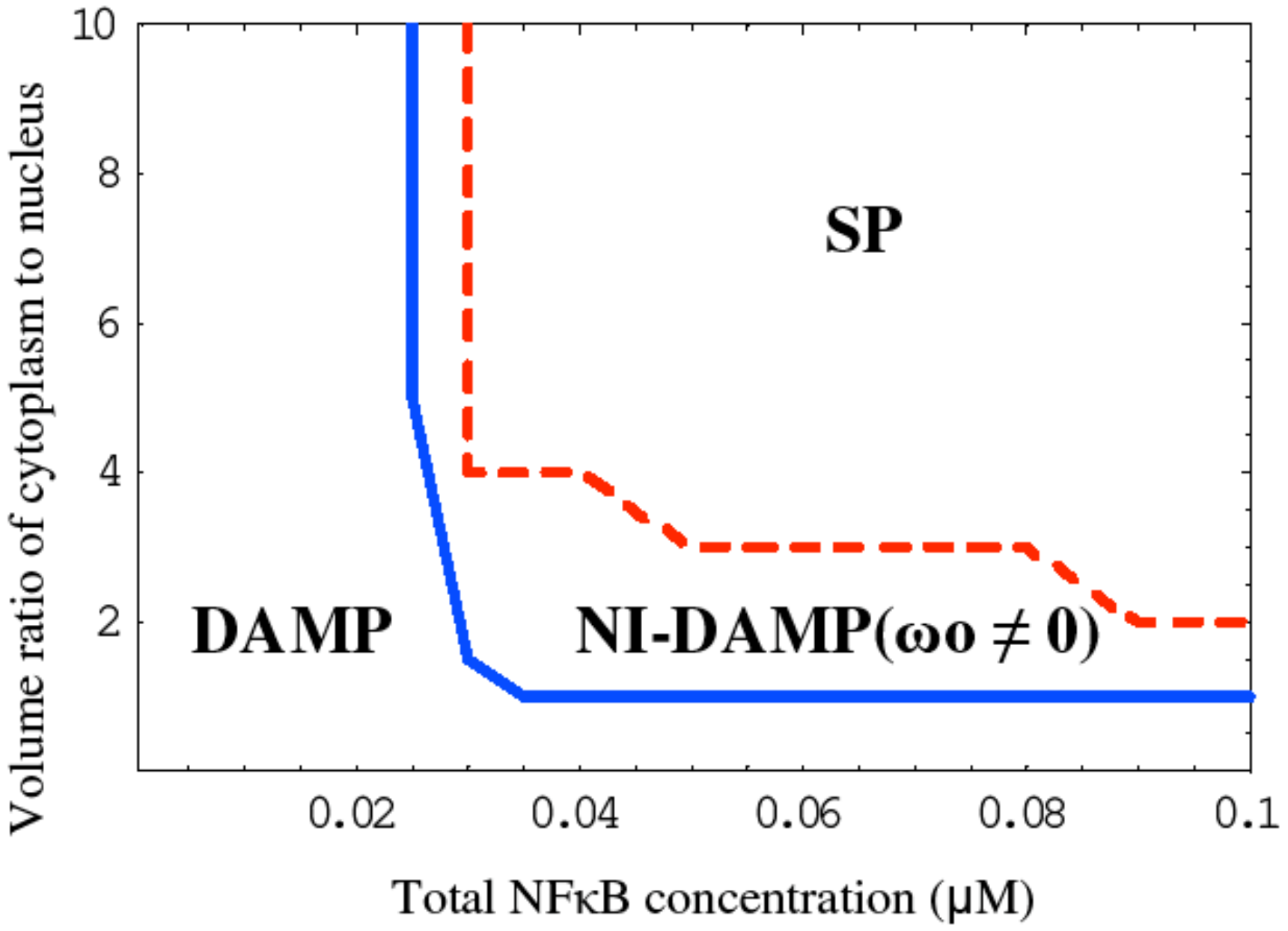

Supporting information figure S2